\documentclass{aa}
\usepackage{savesym}
\usepackage{amsmath}
\savesymbol{iint}
\usepackage{txfonts}
\restoresymbol{TXF}{iint}
\usepackage{graphics,graphicx}
\usepackage{amssymb}
\usepackage{color}
\usepackage[draft]{hyperref}
\usepackage{breakurl}
\usepackage{array}
\usepackage{rotating}
\usepackage{xcolor}
\usepackage{natbib}
\usepackage{float}
\usepackage{flafter}
\usepackage{booktabs}
\usepackage{afterpage}
\usepackage{lipsum}
\usepackage{longtable,pdflscape}
\bibpunct{(}{)}{;}{a}{}{,}

\newcommand{\snra}{MCSNR~J0506$-$7025}
\newcommand{\snrb}{MCSNR~J0527$-$7104}

\newcommand{\chandra}{{\it Chandra}}
\newcommand{\xmm}{{\it XMM-Newton}}
\newcommand{\rosat}{{\it ROSAT}}

\newcommand{\spitzer}{{\it Spitzer}}

\newcommand{\vpshock}{{\tt vpshock}}
\newcommand{\vsedov}{{\tt vsedov}}
\newcommand{\vphabs}{{\tt vphabs}}
\newcommand{\vapec}{{\tt vapec}}
\newcommand{\phabs}{{\tt phabs}}

\newcommand{\vnei}{{\tt vnei}}

\newcommand{\ha}{H$\alpha$}
\newcommand{\sii}{[$\ion{S}{ii}$]}
\newcommand{\oiii}{[$\ion{O}{iii}$]}
\newcommand{\ratio}{[$\ion{S}{ii}$]/H$\alpha$}

\begin{document}

\title{Two evolved supernova remnants with newly identified Fe-rich cores in the Large Magellanic Cloud\thanks{Based on observations obtained with \xmm, an ESA science mission with instruments and contributions directly funded by ESA Member States and NASA}}



\author{P.~J.~Kavanagh \inst{1} \and M.~Sasaki \inst{1} \and L.~M.~Bozzetto \inst{2} \and S.~D.~Points \inst{3} \and E.~J.~Crawford \inst{2} \and J.~Dickel\inst{4} \and \\M.~D.~Filipovi\'c \inst{2} \and F.~Haberl\inst{5} \and P.~Maggi\inst{6} \and E.~T.~Whelan \inst{1}}

 

\institute{Institut f\"{u}r Astronomie und Astrophysik, Kepler Center for Astro and Particle Physics, Eberhard Karls Universit\"{a}t T\"{u}bingen, Sand~1, T\"{u}bingen D-72076, Germany\\ \email{kavanagh@astro.uni-tuebingen.de}
\and Western Sydney University, Locked Bag 1797, Penrith South DC, NSW 1797, Australia
\and Cerro Tololo Inter-American Observatory, Casilla 603, La Serena, Chile
\and Physics and Astronomy Department, University of New Mexico, MSC 07-4220, Albuquerque, NM 87131, USA
\and Max-Planck-Institut f\"{u}r extraterrestrische Physik, Giessenbachstra\ss e, D-85748 Garching, Germany
\and Laboratoire AIM, CEA-IRFU/CNRS/Universit\'e Paris Diderot, Service d'Astrophysique, CEA Saclay, F-91191 Gif sur Yvette Cedex, France}

\date{Received ?? / Accepted ??}

\abstract{}{We present a multi-wavelength analysis of the evolved supernova remnants \object{MCSNR~J0506$-$7025} and \object{MCSNR~J0527$-$7104} in the Large Magellanic Cloud.}{We used observational data from \xmm, the Australian Telescope Compact Array, and the Magellanic Cloud Emission Line Survey to study their broadband emission and used \spitzer\ and \ion{H}{i} data to gain a picture of the environment into which the remnants are expanding. We performed a multi-wavelength morphological study and detailed radio and X-ray spectral analyses to determine their physical characteristics.}{Both remnants were found to have bright X-ray cores, dominated by Fe L-shell emission, consistent with reverse shock heated ejecta with determined Fe masses in agreement with Type~Ia explosion yields. A soft X-ray shell, consistent with swept-up interstellar medium, was observed in \snra, suggestive of a remnant in the Sedov phase. Using the spectral fit results and the Sedov self-similar solution, we estimated the age of \snra\ to be $\sim16-28$~kyr, with an initial explosion energy of $(0.07-0.84)\times10^{51}$~erg. A soft shell was absent in \snrb, with only ejecta emission visible in an extremely elongated morphology extending beyond the optical shell. We suggest that the blast wave has broken out into a low density cavity, allowing the shock heated ejecta to escape. We found that the radio spectral index of \snra\ is consistent with the standard $-0.5$ for SNRs. Radio polarisation at 6~cm indicates a higher degree of polarisation along the western front and at the eastern knot, with a mean fractional polarisation across the remnant of $P \cong (20 \pm 6)\%$. }{The detection of Fe-rich ejecta in the remnants suggests that both resulted from Type~Ia explosions. The newly identified Fe-rich cores in \snra\ and \snrb\ makes them members of the expanding class of evolved Fe-rich remnants in the Magellanic Clouds.}

\titlerunning{Evolved Fe-rich SNRs \snra\ and \snrb}
\maketitle 

\section{Introduction}
Supernovae are powerful stellar explosions which are important in many fields of astrophysics. They are key components of the engine which drives the physical and chemical evolution of the interstellar medium (ISM) in galaxies and are sources of cosmic rays \citep[for a review, see][]{Vink2012}. There are two types of supernovae. Core-collapse explosions signal the death of massive, short-lived stars ($\gtrsim$8~M$_{\sun}$). Type Ia supernovae are thought to result from the disruption or explosion of a carbon-oxygen (C-O) white dwarf surpassing the Chandrasekhar limit, either via accretion from a low mass stellar companion (the single degenerate scenario), or through the merger of two white-dwarfs (the double degenerate scenario). However, it has been recently shown that Type Ia SNe may also result from the explosion of sub-Chandrasekhar white dwarfs \citep[e.g.,][]{Sim2010,Woosley2011}. It is still unclear as to which of the explosion channels is dominant in the Universe \citep[][and references therein]{Maoz2008}. 
\par Following the explosion, the freshly produced nucleosynthesis products are blown outward, seeding the ISM of the host galaxy. The interaction of these expanding ejecta with the ambient medium creates a lasting imprint in the form of a supernova remnant (SNR). Recently, a new class of evolved remnants has been discovered in the Magellanic Clouds, the defining characteristic of which is the presence of an X-ray bright Fe-rich core, consistent with reverse shock-heated Type~Ia ejecta \citep{Borkowski2006}. An additional feature of the Fe-rich plasma is that the gas is close to or in collisional ionisation equilibrium (CIE), i.e., the post-shock ionisation balance has been restored. This presents a problem for standard Type~Ia models \citep[e.g.,][]{Dwarkadas1998,Badenes2003,Badenes2005} which predict that the central Fe-rich gas should be of low density and the ionisation timescales short, contrary to what is observed. \citet{Borkowski2006} argue that, to explain the higher than expected central densities, some pre-explosion seeding of the circumstellar medium (CSM) by the stellar progenitor must have occurred, with the best candidates being the so-called `prompt' single degenerate systems \citep{Mannucci2006,Sullivan2006,Aubourg2008}. These `prompt' events are produced by intermediate mass stars just below the core-collapse explosion limit \citep[$\sim$3.5--8~M$_{\sun}$,][]{Aubourg2008} and are capable of enhancing the CSM density via ejection of the stellar envelope prior to the SN event. As stated by \citet{Borkowski2006}, at later stages of the resulting SNR evolution, the Fe ejecta would manifest as a dense Fe-rich gas close to the centre of the SNR.
\par Nine of these objects have been detected in the \object{Large Magellanic Cloud} (LMC) so far: \object{0454--67.2} \citep{Seward2006}; \object{0548--70.4} and \object{0534--69.9} \citep{Hendrick2003}; \object{DEM~L316A} \citep{Nish2001,Williams2005}; \object{DEM~L238} and \object{DEM~L249} \citep{Borkowski2006}; \object{MCSNR~J0508--6902} \citep{Bozzetto2014}; \object{MCSNR~J0508--6830} and \object{MCSNR~J0511--6759} \citep{Maggi2014}. In this paper we report on new X-ray observations and multi-wavelength studies of two known SNRs in the LMC which we have found to exhibit bright Fe-rich interiors and are therefore new additions to this relatively recent class of evolved SNRs.
\par \citet{Davies1976} used optical and radio data to identify many SNRs and candidate SNRs in the Magellanic Clouds. One of their candidate SNRs, LMC~B0507$-$7029, was later confirmed by \citet{Filipovic1998} through cross-correlation of LMC radio and X-ray source catalogues. \citet{Haberl1999} reported on \rosat\ Position Sensitive Proportional Counter (PSPC) observations of the LMC, labelling this source  \object{[HP99]~1139}. While the PSPC source was confirmed as extended, its large off-axis angle of $\sim49\arcmin$ prevented a reliable SNR classification. \citet{Payne2008} later confirmed the SNR classification of this object using long-slit optical spectroscopy of the bright filaments. We hereafter refer to this object as \snra, following the Magellanic Cloud Emission Line Database\footnote{\burl{http://www.mcsnr.org/}} (MCSNR) nomenclature, and the determined position from our new X-ray data (see Section~\ref{1139-mwm}).
\par \citet{Haberl1999} also identified the \rosat\ PSPC source \object{[HP99]~1234}, located next to the \object{N~206} \ion{H}{ii} complex \citep{Henize1956} in the LMC. It was classified as a candidate SNR based on its X-ray hardness ratios and extent. An optical SNR candidate was identified at the location of [HP99]~1234 by \citet{Kavanagh2012} in a study of the N~206 superbubble using data from the Magellanic Cloud Emission Line Survey \citep[MCELS,][]{Smith2006}, as shown in their Fig.~7. A more detailed analysis of this object using the \rosat\ and MCELS data, supplemented with Molonglo Observatory Synthesis Telescope (MOST) radio continuum data, was performed by \citet[][hereafter KSP13]{Kavanagh2013}, who confirmed its SNR nature. KSP13 suggested that \snrb\ is most likely in the Sedov phase of its evolution, determining an age of $\sim25$~kyr, though this is based on a somewhat uncertain ambient density estimate. Interestingly, only very faint 36~cm emission was detected from the remnant, with the emission well below the detection limit at all other frequencies, making it one of the weakest SNRs ever detected in the radio regime. Using the local stellar population and star formation history, KSP13 suggested a Type~Ia explosion scenario for \snrb, though a CC explosion due to a star `kicked' from N~206 could not be ruled out.
\par Our work on the multi-frequency studies of \snra\ and \snrb\ using X-ray, radio, \ion{H}{i}, IR, and optical emission line data is arranged as follows: the observations and data reduction are described in Section~2;  the radio analysis of \snra\ and X-ray analysis of both remnants are outlined in Sections~3 and 4, respectively; results are given and discussed in Section~5; and finally we summarise our work in Section~6.

\section{Observations and data reduction}
\label{odr}
\subsection{Optical}
\begin{figure*}[!t]
\begin{center}
\resizebox{\hsize}{!}{\includegraphics[trim= 0cm 0cm 0cm 0cm, clip=true, angle=0]{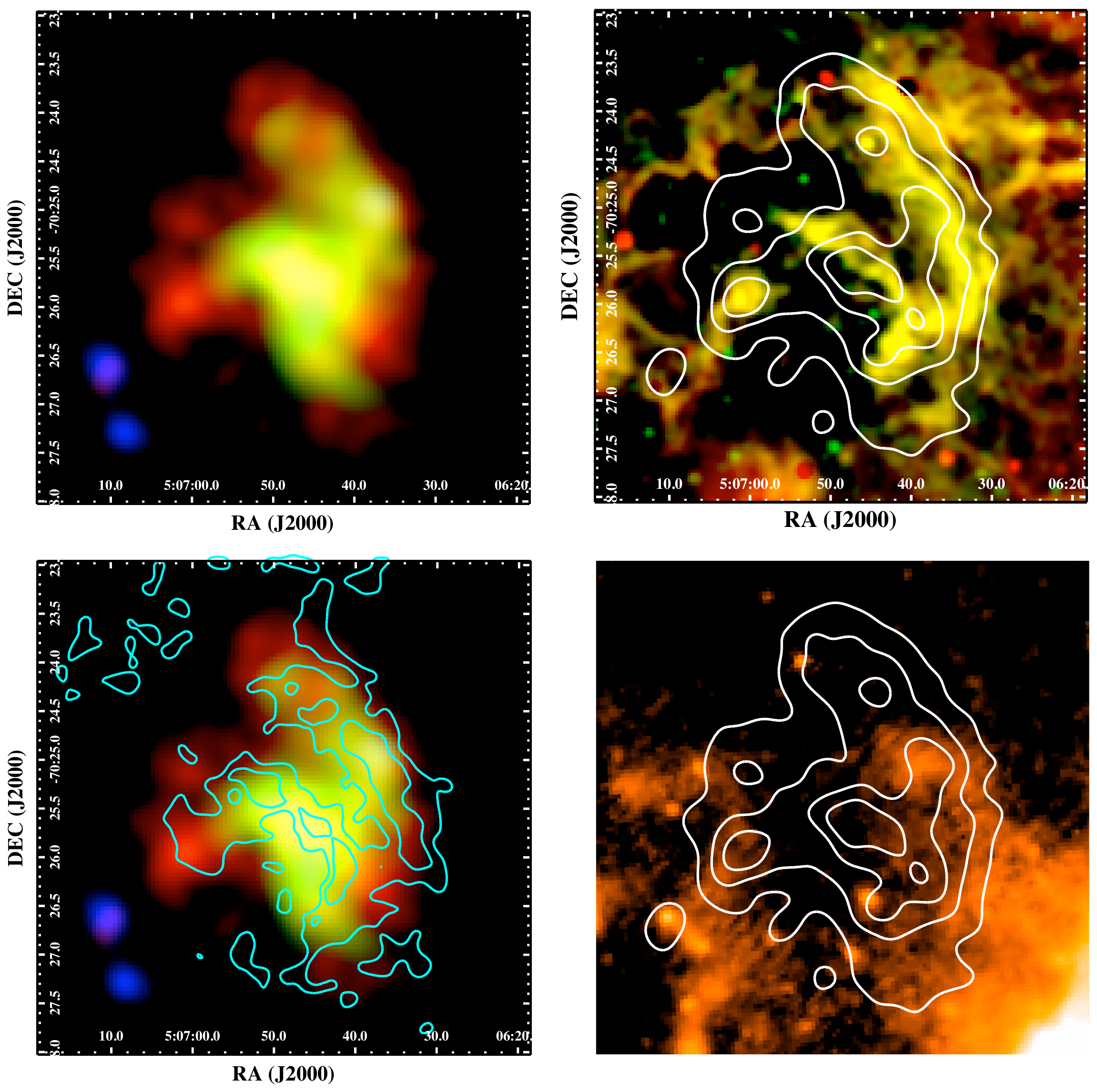}}
\caption{{\bf Top left:} \xmm\ EPIC image of \snra\ in false colour with RGB corresponding to 0.3--0.7~keV, 0.7--1.1~keV, and 1.1--4.2~keV, respectively.  {\bf Top right:} Continuum subtracted MCELS image of \snra\ with \ha\ in red and \sii\ in green overlaid with 0.3--0.7~keV contours. The lowest contour level represents 3$\sigma$ above the average background surface brightness, with the remaining levels marking 25\%, 50\%, and 75\% of the maximum above this level.  {\bf Bottom left:} Same as top left but with \ratio\ contours with the lowest level corresponding to \sii/\ha~=~0.67, and the remaining levels at 25\%, 50\%, and 75\% of the maximum above this level. {\bf Bottom right:} \spitzer\ MIPS 24~$\mu$m image of the \snra\ region with soft X-ray contours from top right. The image scale is the same as in all other panels.}
\label{1139_4panel}
\end{center}
\end{figure*}

\begin{figure*}[!t]
\begin{center}
\resizebox{\hsize}{!}{\includegraphics[trim= 0cm 0cm 0cm 0cm, clip=true, angle=0]{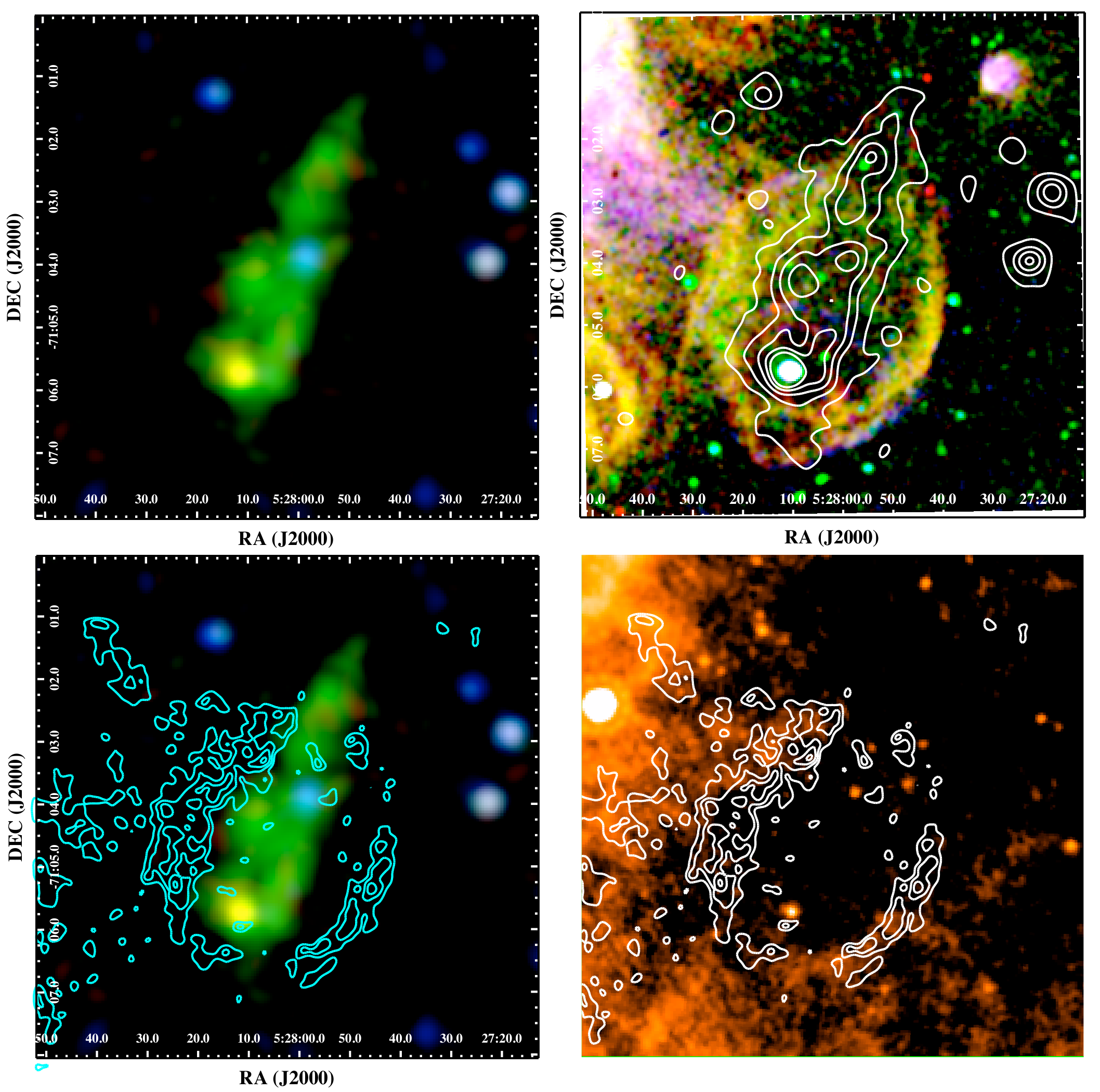}}
\caption{{\bf Top left:} \xmm\ EPIC image of \snrb\ in false colour with RGB corresponding to 0.3--0.7~keV, 0.7--1.1~keV, and 1.1--4.2~keV, respectively.  {\bf Top right:} Continuum subtracted MCELS image of \snrb\ from KSP13 with \ha\ in red, \sii\ in green, and \oiii\ in blue overlaid with 0.3--0.7~keV contours. The lowest contour level represents 3$\sigma$ above the average background surface brightness, with the remaining levels marking 25\%, 50\%, and 75\% of the maximum above this level.  {\bf Bottom left:} Same as top left but with \ratio\ contours with the lowest level corresponding to \sii/\ha~=~0.67, and the remaining levels at 25\%, 50\%, and 75\% of the maximum above this level (from KSP13). {\bf Bottom right:} \spitzer\ MIPS 24~$\mu$m image of the \snrb\ region with \ratio\ contours from bottom left . The image scale is the same as in all other panels.}
\label{1234_4panel}
\end{center}
\end{figure*}

The MCELS observations \citep{Smith2006} were taken with the 0.6 m University of Michigan/Cerro Tololo Inter-American Observatory (CTIO) Curtis Schmidt Telescope equipped with a SITE 2048 $\times$ 2048 CCD, producing individual images of $1.35^{\circ} \times 1.35^{\circ}$ at a scale of 2.3$\arcsec$ pixel$^{-1}$. The survey mapped both the LMC ($8^{\circ} \times 8^{\circ}$) and the Small Magellanic Cloud ($3.5^{\circ} \times 4.5^{\circ}$) in narrow bands covering [\ion{O}{iii}]$\lambda$5007 \AA, H$\alpha$, and [\ion{S}{ii}]$\lambda$6716, 6731 \AA, in addition to matched green and red continuum bands. The survey data were flux calibrated and combined to produce mosaicked images. We extracted cutouts centred on \snra\ from the MCELS mosaics. We subtracted the continuum images from the corresponding emission line images, thereby removing the stellar continuum and revealing the full extent of the faint diffuse emission. Finally, we divided the continuum subtracted \sii\ image by the continuum subtracted \ha\ image to get a \ratio\ map of \snra, with regions of \ratio~$>0.4$ indicative of the presence of an SNR, or \ratio~$>0.67$ in the case of an SNR emission being contaminated by photoionisation from a nearby \ion{H}{ii} region \citep{Mathewson1973,Fesen1985}, the latter being the case for \snra. We note that the MCELS observations of \snrb\ have already been reported in KSP13, though we use the images again here to aid in the multi-wavelength morphology analysis. The continuum subtracted emission line images for \snra\ and \snrb\ are shown in Figs.~\ref{1139_4panel}-top right and \ref{1234_4panel}-top right, respectively.

\subsection{Radio}
\label{radio-obs}
\begin{figure}
\resizebox{\hsize}{!}{\includegraphics[trim=0 0 0 0, clip]{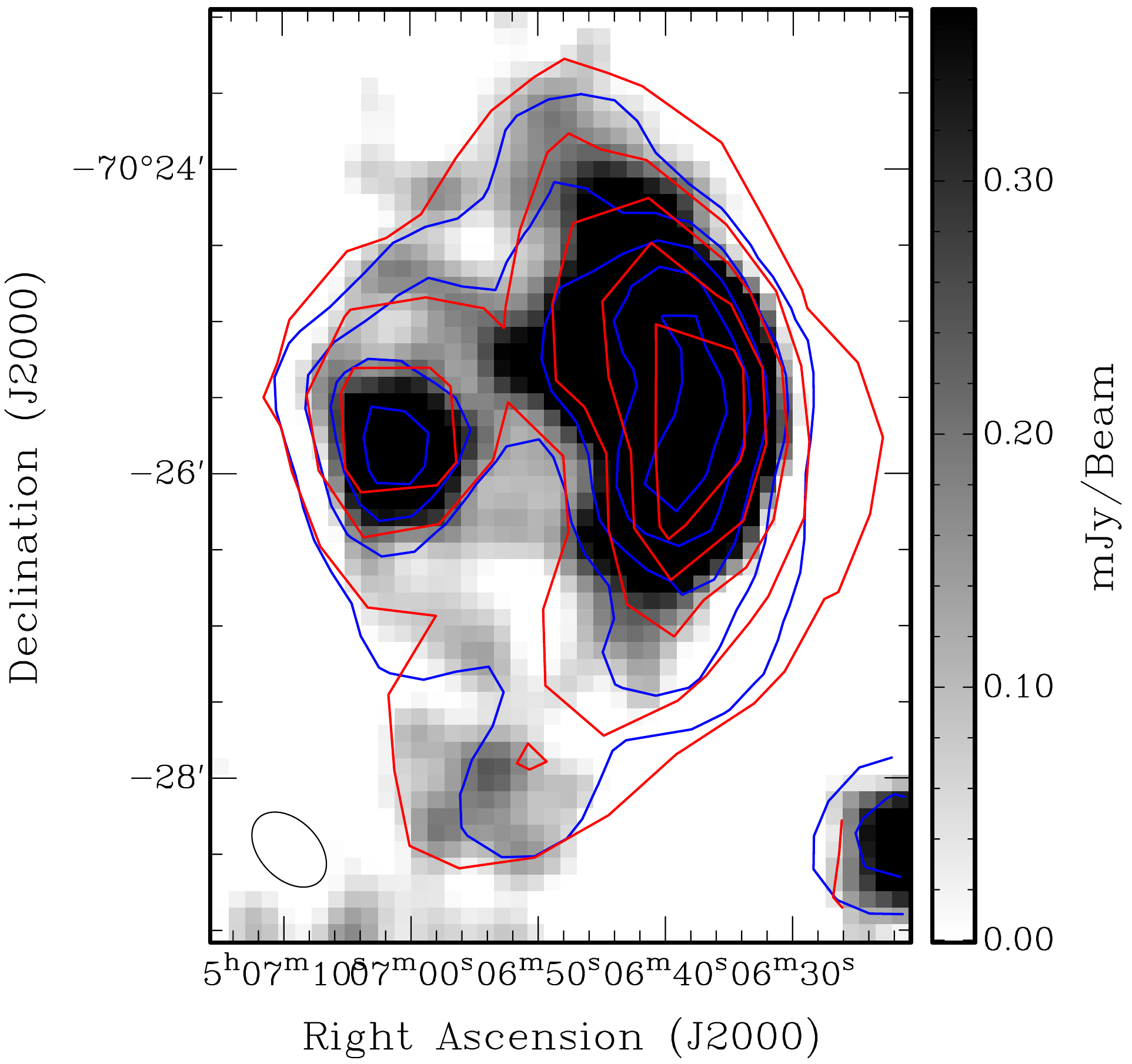}}
\caption{ATCA 6~cm intensity image of \snra\ overlaid with 36~cm contours (red; 1 to 11~mJy in steps of 2~mJy) and 20~cm contours (blue; 1.8 to 6.6~mJy in steps of 1.2~mJy). The ellipse in the lower left corner represents the synthesised beamwidth of 34.7~$\times$~23.1\arcsec, and the sidebar shows the flux density scale and is given in mJy/Beam. \label{radioemission}}
\end{figure}

We observed \snra\ with the Australian Telescope Compact Array (ATCA) on November 15 and 16 2011, using the new Compact Array Broadband Backend (CABB). The ATCA array configuration EW367 was used and observations were taken simultaneously at $\lambda = 3$ and 6~cm ($\nu$=9\,000 and 5\,500~MHz) using the dual-frequency mode. Baselines formed with the $6^\mathrm{th}$ ATCA antenna were excluded as the other five antennas were arranged in a compact configuration. The observations were carried out in the so called ``snap-shot'' mode, totalling $\sim$64 minutes of integration over a 14 hour period. PKS~B1934-638 was used for flux density calibration\footnote{Flux densities were assumed to be 5.098~Jy at 6~cm and 2.736~Jy at 3~cm.}  and PKS~B0530-727 was used for secondary (phase) calibration. The phase calibrator was observed twice every hour for a total 78 minutes over the whole observing session. The \texttt{miriad}\footnote{http://www.atnf.csiro.au/computing/software/miriad/}  \citep{1995ASPC...77..433S} and \texttt{karma}\footnote{http://www.atnf.csiro.au/computing/software/karma/} \citep{1995ASPC...77..144G} software packages were used for reduction and analysis. The 6~cm image (Fig.~\ref{radioemission}) has a resolution of $34\farcs7\times23\farcs1$ at PA=44.4\degr, and an estimated rms noise of 0.2~mJy/beam.  Similarly, we made an image of \snra\ at 3~cm, which has a resolution of $23\farcs7\times15\farcs3$ at PA=44.1\degr, and an estimated rms noise of 0.1~mJy/beam.

In addition to these observations, we also make use of a 36~cm (843~MHz) MOST mosaic image \citep[as described in][]{1984AuJPh..37..321M}, a 20~cm (1\,377~MHz) mosaic image from \citet{2007MNRAS.382..543H} (the emission at both of the frequencies can be seen as superimposed contours in Fig.~\ref{radioemission}), and 6~cm and 3~cm mosaics \citep{2010AJ....140.1567D}.

The analysis of radio emission from \snrb\ has already been reported in KSP13.

\subsection{X-ray}
\snra\ and \snrb\ were observed by \xmm\ \citep{Jansen2001} on May 3 2013 (Obs. IDs 0720440201, PI M. Sasaki) and May~31~-~June~1 2014 (Obs. IDs 0741800101, PI P.J. Kavanagh), respectively. The primary instrument for the observations was the European Photon Imaging Camera (EPIC), which consists of a pn CCD \citep{Struder2001} and two MOS CCD imaging spectrometers \citep{Turner2001}. The observational data were reduced using the standard reduction tasks of SAS\footnote{Science Analysis Software, see http://xmm.esac.esa.int/sas/} version 14.0.0, filtering for periods of high particle background. This resulted in $\sim11$~ks for the EPIC-pn and $\sim28$~ks each for the EPIC-MOS detectors for \snra, and $\sim27$~ks for the EPIC-pn and $\sim31$~ks each for the EPIC-MOS detectors for \snrb, which were available for further analysis.

\subsection{Archival data}
\subsubsection{Infrared}
\label{ir}
The cold environment surrounding the remnants can be revealed by infrared (IR) emission. To aid in the discussion of the morphology and environment of the remnant, we made use of data from the SAGE survey of the LMC \citep{Meixner2006} with the \spitzer\ \textit{Space Telescope} \citep{Werner2004}. During the SAGE survey, a $7\degr\times7\degr$ area of the LMC was observed with the Infrared Array Camera \citep[IRAC,][]{Fazio2004} in the 3.6~$\mu$m, 4.5~$\mu$m, 5.8~$\mu$m, and 8~$\mu$m bands, and with the Multiband Imaging Photometer \citep[MIPS,][]{Rieke2004} in the 24~$\mu$m, 70~$\mu$m, and 160~$\mu$m bands. The MIPS 24~$\mu$m images provide us with a picture of the stochastically and thermally heated dust in the region of \snra\ and \snrb, with spatial resolution comparable to \xmm, to give an indication of the distribution of cool material. We obtained the 24~$\mu$m MIPS mosaicked, flux-calibrated (in units of MJy~sr$^{-1}$) images processed by the SAGE team from the NASA/IPAC Infrared Science Archive\footnote{See \burl{http://irsa.ipac.caltech.edu/data/SPITZER/SAGE/}}. The pixel sizes correspond to $4.8\arcsec$ for the 24~$\mu$m band, $\sim1.2$~pc at the LMC distance.

\subsubsection{\ion{H}{i}}
\label{hi}
To gain an understanding of the neutral H in the regions of \snra\ and \snrb\ we used data from the \ion{H}{i} survey of the LMC by the ATCA and Parkes facilities\footnote{\burl http://www.atnf.csiro.au/research/HI/mc/queryForm.html}, which are described in detail by \citet{Stav2003} and \citet{Kim2003}. 

\section{Radio analysis of \snra}
\label{raan}
\subsection{Radio spectral index}
\begin{table*}
\begin{center}
 \caption{Spectral index estimates of \snra. Column (1) lists the region of interest. Columns (2) and (3) show the spectral indices from taking direct integrated flux density measurements. Columns (4), (5), and (6) show the results of T-T plots to the listed images. Column (7) shows the median value of the spectral indices taken from a spectral map at a resolution of 20~cm, while column (8) is the same apart from the resolution being at 36~cm, comparable to the same resolution of the T-T plots. In all columns related to spectral index values, the wavelengths of the images used are given in brackets.}
 \label{tbl-1}
 \begin{tabular}{@{}cccccccccc}
 \hline
  		& $\alpha_S$	&	$\alpha_S$ 	&	$\alpha_{TT}$	&	$\alpha_{TT}$	&	$\alpha_{TT}$  &	$\alpha_{S_{med}}$  \\
  	 		& (36-20-6-3~cm)	&	(20-6-3~cm)	&		(36-20~cm)	&		(36-6~cm)	&	(20-6~cm)		&		(20-6~cm)			\\
 \hline\
All	&$-0.25\pm0.02$&$-0.28\pm0.02$	&$-0.64$			&$-0.47$				&$-0.36$			&		$-0.33$		\\
 \hline
 \end{tabular}
 \end{center}
 \end{table*}

To estimate the spectral index of \snra, three different methods were used. The first is a simple fit to the total integrated flux densities taken at the different wavelengths. The second, $\alpha_{TT}$, uses the temperature - temperature (T-T) plot method which has the benefit that it is immune to the zero-level in the individual maps. The third, $\alpha_{S_{med}}$, takes the median value of a spectral index map. For all three methods, unless otherwise stated, the data at each wavelength were convolved down to the lowest resolution (36~cm; beam size = $46\farcs4\times43\farcs0$) to make the images comparable.  

\begin{figure*}
\resizebox{\hsize}{!}{\includegraphics[trim=0 0 0 0, clip]{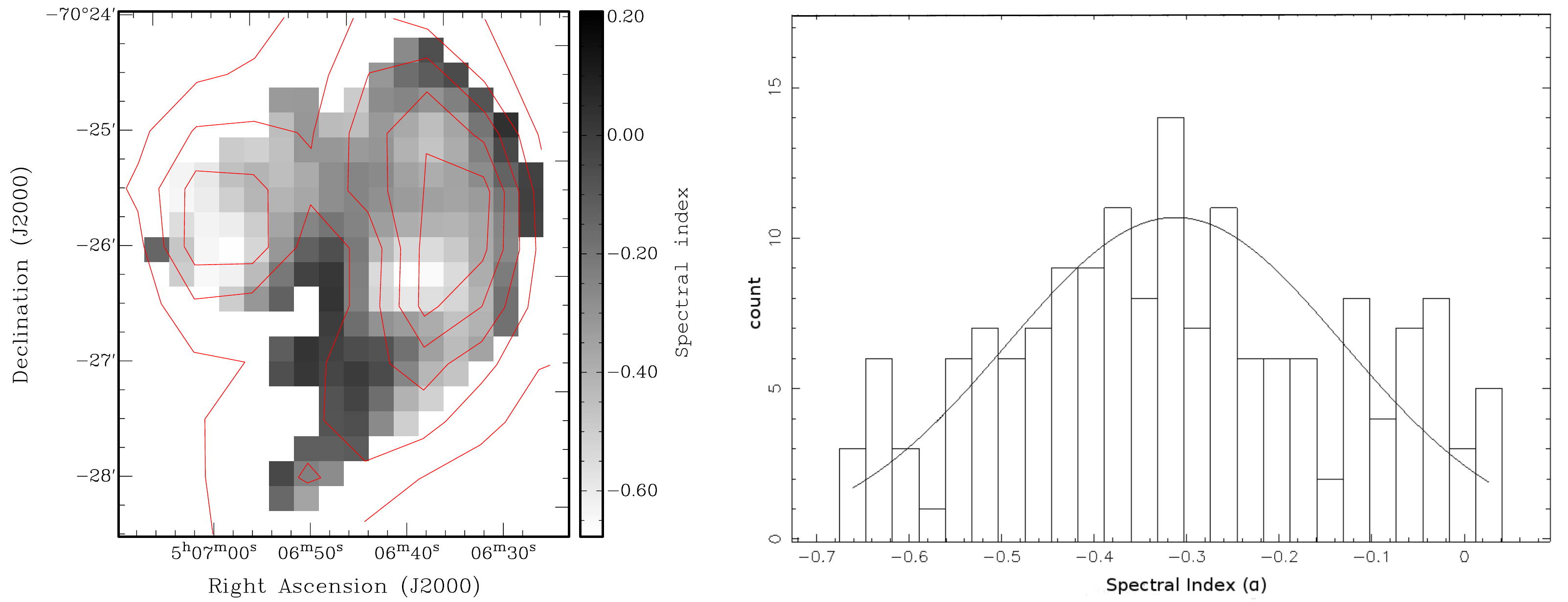}}
\caption{{\bf Left:} Spectral index map between 20 and 6~cm for \snra. The contours are the same 36~cm contours as shown in Fig.~\ref{radioemission}. {\bf Right:} Histogram of the distribution of spectral index values from the spectral map in left.
 \label{spchist}}
\end{figure*}

\begin{figure}[!h]
\resizebox{\hsize}{!}{\includegraphics[trim=0 0 0 0, clip]{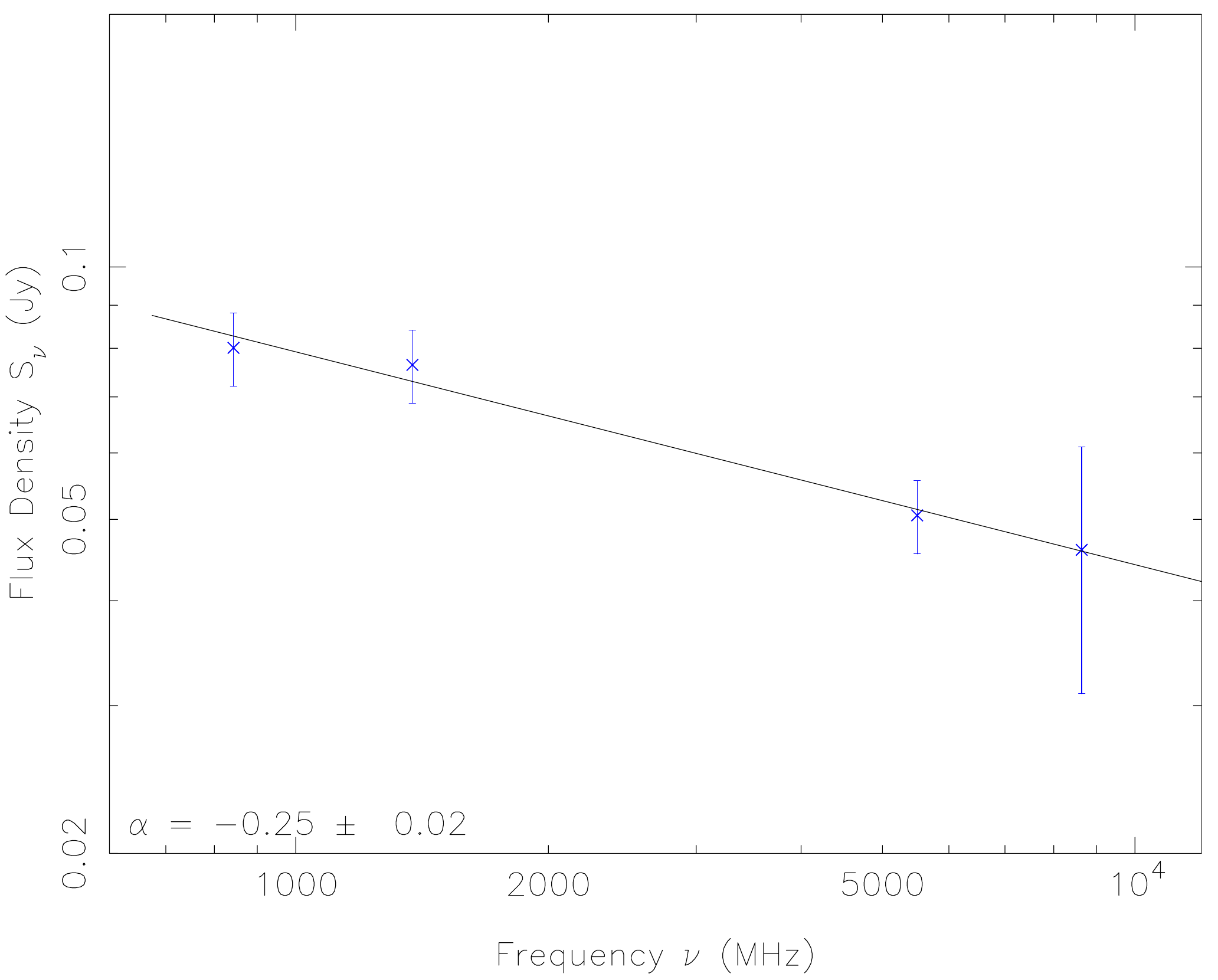}}
\caption{Spectral index fit to \snra\ using integrated flux density measurements at 36, 20, 6, and 3~cm. 
 \label{spcidx}}
\end{figure}

Integrated flux density measurements were taken at 36~cm (80.1~mJy), 20~cm (76.4~mJy), and 6~cm (50.6~mJy). Uncertainties in these flux density measurements predominately arose from defining the edge of the remnant. We estimate that the errors in all flux density estimates are within ~10\%. The integrated flux density measurement at 3~cm is a little more problematic due to contamination from the faint \ion{H}{ii} region to the west of the SNR. We estimated the flux density at 3~cm to be 46~($\pm$15)~mJy. The resulting plot and fit are given in Fig.~\ref{spcidx}, where $\alpha_{S} = -0.25\pm0.02$. This is relatively flat compared to the typical SNR spectral index of $\sim-0.5$.


To verify the spectral index of the emission, a spectral map was created only between 20 and 6~cm wavelengths. The 36~cm image was not used in this scenario as the missing short spacings responsible for the large scale emission severely impacted the flux density of the image, resulting in non-sensical values of $\alpha$. The spectral index map was created by reprocessing both sets of data to a common \textit{u~--~v} range, and then fitting $S \propto \nu^{\alpha}$ pixel by pixel using both images, and is shown in Fig.~\ref{spchist}-left. The resulting spectral pixel values were binned, and a histogram of their distribution is shown in Fig.~\ref{spchist}-right. The 20 and 6~cm images were masked at levels of 2.0 and 1.8~mJy, respectively. Values that fell below these cutoffs were blanked from the output image. The resulting median spectral index from this data is $\alpha_{S_{med}} = -0.33$.

\begin{figure}[!h]
\resizebox{\hsize}{!}{\includegraphics[trim=0 0 0 0, clip]{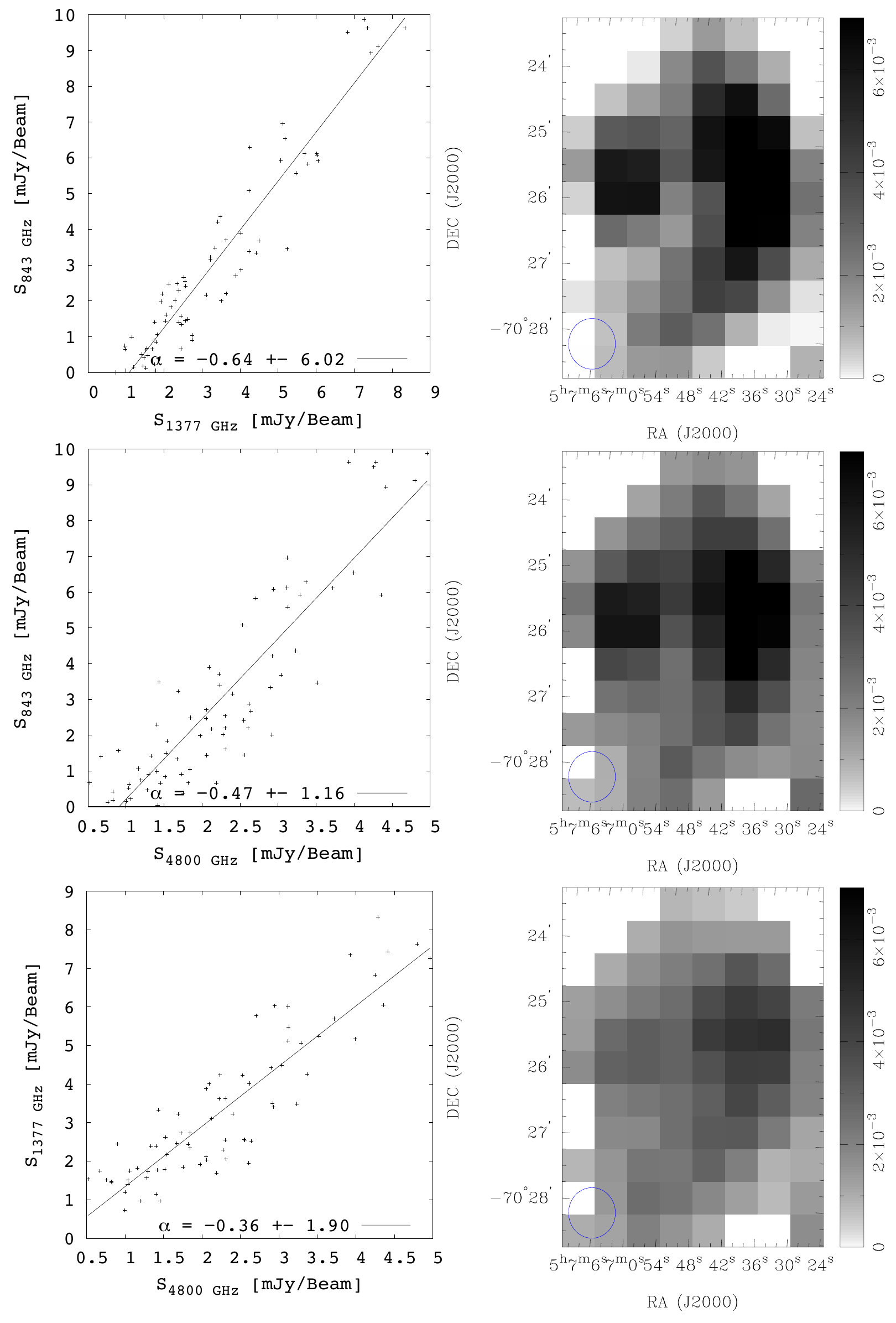}}
\caption{{\bf Left:} T-T plots of \snra\ between 36~--~20~cm, 36~--~6~cm, and 20~--~6 cm (top to bottom). {\bf Right:} Intensity images showing the region and strength of the emission from 36, 20, and 6~cm images (top-to-bottom).
 \label{tt}}
\end{figure}

We also estimated the spectral index using the T-T plot method \citep{Costain1960,Turtle1962}, i.e., the spectral index is determined by finding the scaling factor which gives the best agreement between temperature curves at given frequencies. We created T-T plots between 36 -- 20~cm, 36 -- 6~cm, and 20 -- 6 cm (see Fig.~\ref{tt}). This method is seen as the most reliable as it neglects any external flux (e.g., background emission). The resulting spectral indices, $-0.64, -0.47, -0.36$, show the non-thermal nature of the emission. Taking the mean spectral index of the three plots results in $\alpha=-0.49$, well in line with the index characteristic of SNRs of $-0.5$.  \\

\noindent Although there is clearly some variation in values found from the various methods and images used, all the spectral index results point to the emission being non-thermal in nature. The most reliable results, the T-T plot method, showed a spectral index inline with the standard $-0.5$. If we do not take into account the other results (i.e., the flatter spectra from analyses such as the spectral map between 20 and 6~cm), we are left with a spectral index that is characteristic of an evolved remnant.

\subsection{Magnetic field}
\begin{figure}
\resizebox{\hsize}{!}{\includegraphics[trim=0 0 0 0, clip]{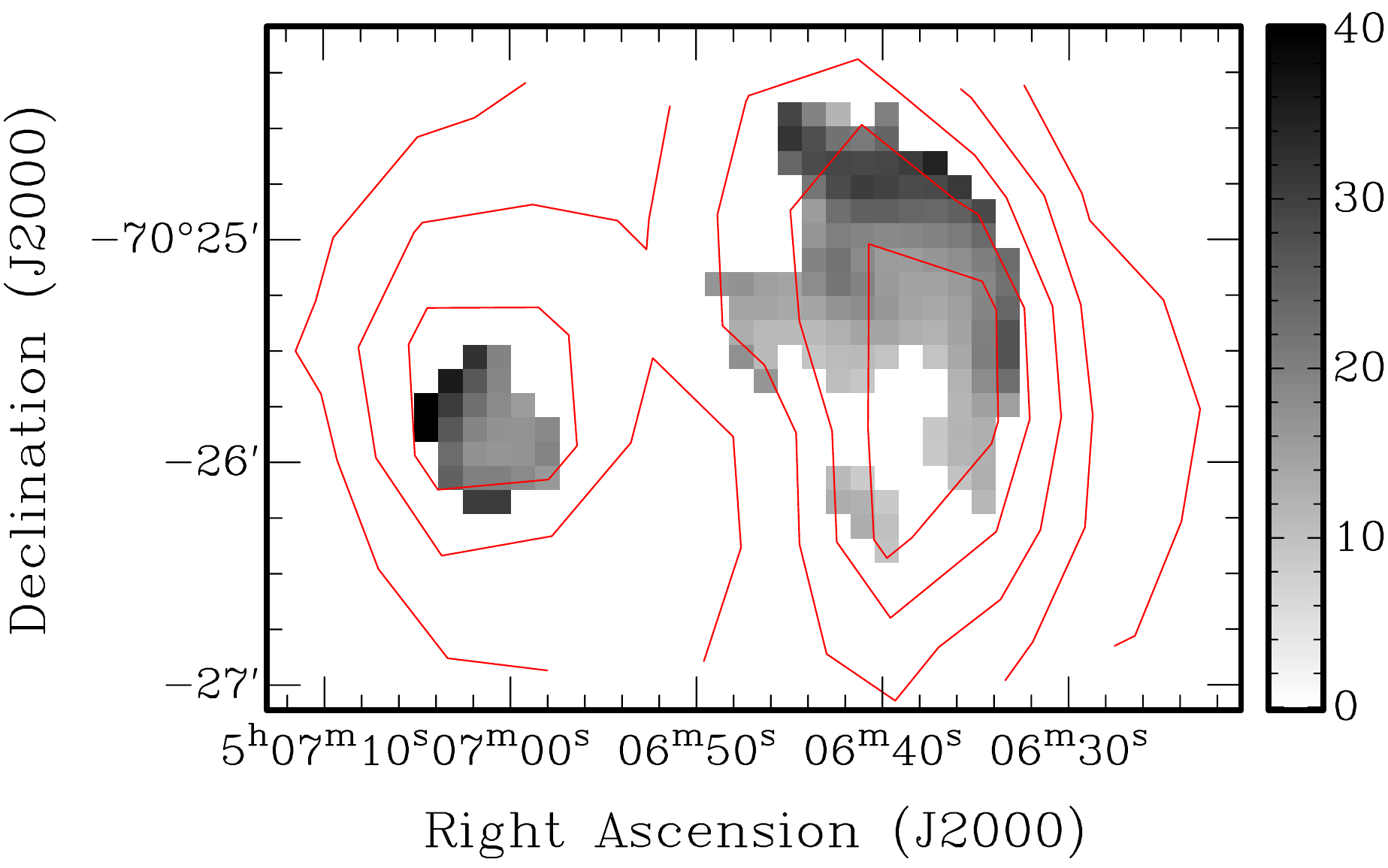}}
\caption{Fractional polarisation ($p$) of \snra\ at 6~cm. The side bar quantifies the level of fractional polarisation in percent. Pixels where the error was half the measured fractional polarisation or greater were blanked from the image. The contours are the same 36~cm contours as shown in Fig.~\ref{radioemission}.
 \label{polm_masked}}
\end{figure}

A polarisation image of \snra\ was created at 6~cm using the \texttt{miriad} task \texttt{impol}, from \textit{Q} and \textit{U} Stokes parameters. The mean fractional polarisation was calculated using flux density and polarisation:

\begin{equation}
P=\frac{\sqrt{S_{Q}^{2}+S_{U}^{2}}}{S_{I}}
\end{equation}

\noindent where $S_{Q}, S_{U}$, and $S_{I}$ are integrated intensities for the \textit{Q}, \textit{U}, and \textit{I} Stokes parameters. A signal-to-noise cut-off of 2$\sigma$ was used for the \textit{Q} and \textit{U} images, and a level of 8$\sigma$ for the intensity image. Values that fell below these cut-off levels are blanked in the output image. Additionally, the fractional polarisation map \textit{p} (shown in Fig.~\ref{polm_masked}) was compared with its associated error map $\sigma_p$, and pixels were blanked for $p/\sigma_p\leq2$. The emission in the west shows clear amplification of the magnetic field toward the shock front, with values of $P \approx 30\%$. Similarly, the emission in the east shows this amplification, where fractional polarisation values reach as high as 45\%. 

\begin{figure}
\resizebox{\hsize}{!}{\includegraphics[trim=0 0 0 0, clip]{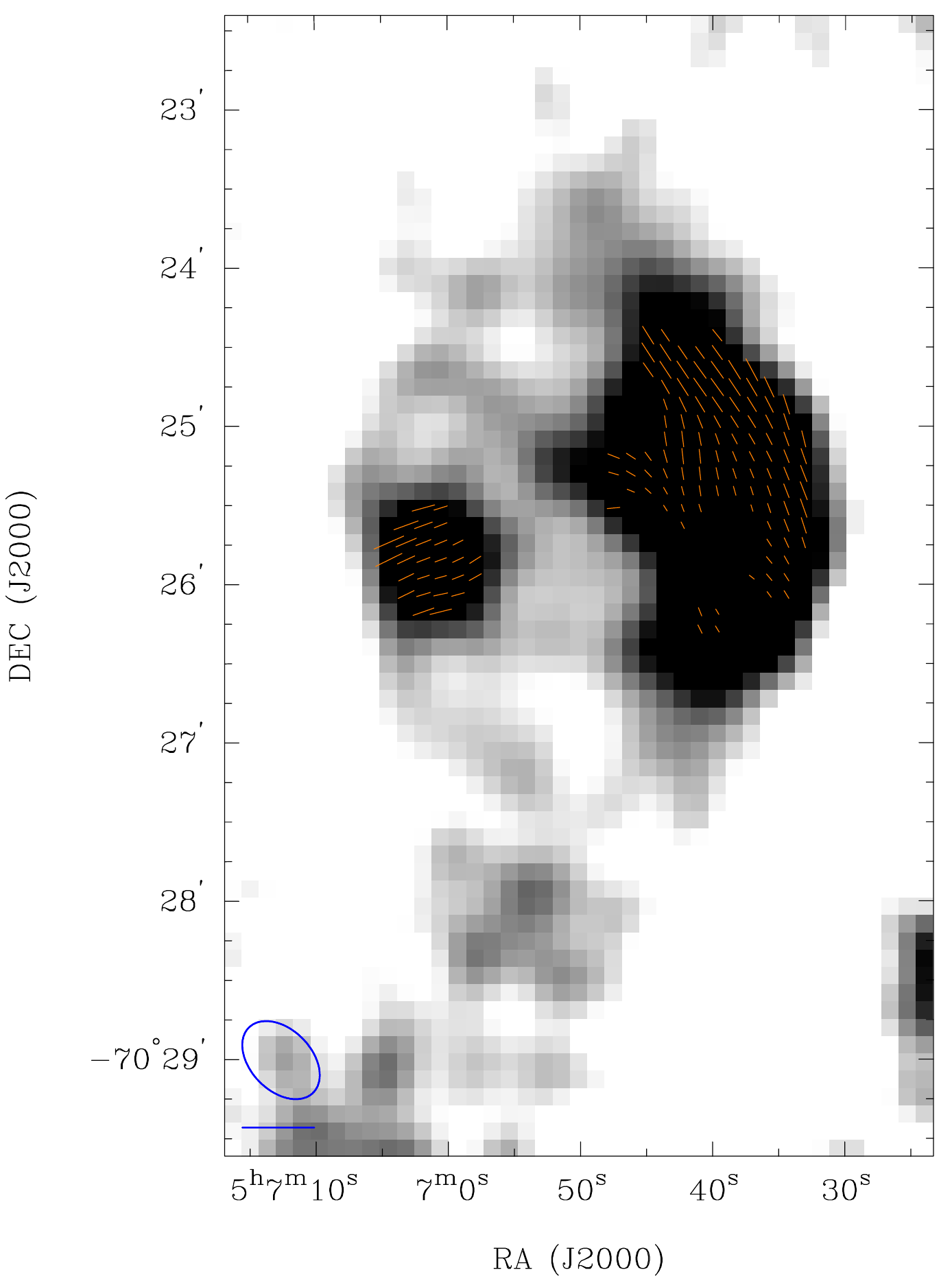}}
\caption{Magnetic field line vectors of \snra\ from the ATCA 6~cm observations overlaid on an intensity image at the same wavelength. The ellipse in the lower left corner represents the synthesised beamwidth of 34.7~$\times$~23.1\arcsec.
 \label{6cmpol}}
\end{figure}

Mean polarisation across the remnant at 6~cm was found to be $P \cong (20 \pm 6)\%$. As there are two distinct regions of this SNR, we have taken independent measurements for both. We find a mean polarisation of $P \cong (19 \pm 6)\%$ for the SNR emission to the west, and a slightly higher $P \cong (24 \pm 7)\%$ for the emission in the east. The orientation of the magnetic field vectors can be seen in Fig.~\ref{6cmpol}. The length of the vectors have been halved for display purposes. 

In comparison to other LMC SNRs arising from a Type~Ia SNe, \snra\ shows similar mean levels of fractional polarisation: \object{MCSNR~J0509--6731} with 26$\pm$13\% at 6~cm \citep{2014MNRAS.440.3220B}; \object{MCSNR~J0533--7202} with 12$\pm$7\% \citep{2013MNRAS.432.2177B}; \object{MCSNR~J0536--7038} with 35$\pm$8\% \citep{2014Ap&SS.351..207B}; while \object{MCSNR~J0519--6902} exhibited a much lower mean fractional polarisation at a level of 2.2 and 3.2\% at 6 and 3~cm, respectively \citep{2012SerAJ.185...25B}.

\section{X-ray analysis}
\label{xan}
\subsection{X-ray imaging}
\label{x-ray-imaging}
We produced images and exposure maps in various energy bands from the flare-filtered event lists for each EPIC instrument in each observation. We filtered for single and double-pixel events (\texttt{PATTERN} $\lesssim4$) from the EPIC-pn detector, with only single pixel events considered below 0.5 keV to avoid the higher detector noise contribution from the double-pixel events at these energies. All single to quadruple-pixel events (\texttt{PATTERN} $\lesssim12$) were considered for the MOS detectors. 

\par We used three energy bands suited to the analysis of the spectra of SNRs. A soft band from 0.3--0.7 keV includes strong lines from O; a medium band from 0.7--1.1 keV comprises Fe L-shell lines as well as Ne He$\alpha$ and Ly$\alpha$ lines; and a hard band (1.1--4.2 keV) which includes lines from Mg, Si, S, Ca, and Ar.

\par We subtracted the detector background from the images using filter-wheel-closed data (FWC). The contribution of the detector background to each EPIC detector was estimated from the count rates in the corners of the images, which were not exposed to the sky. We then subtracted appropriately-scaled FWC data from the raw images. We merged the EPIC-pn and EPIC-MOS images from each observation into combined EPIC images and performed adaptive smoothing of each using a template determined from the combined energy band (0.3--4.2~keV) EPIC image. The sizes of Gaussian kernels were computed at each position in order to reach a signal-to-noise ratio of five, setting the minimum full width at half maximum of the kernels to $\sim8\arcsec$. In the end the smoothed images were divided by the corresponding vignetted exposure maps. Finally, we produced three-colour images of \snra\ and \snrb\, which are shown in Figs. \ref{1139_4panel}-top left and \ref{1234_4panel}-top left, respectively.

\subsection{Point sources}
\label{src_proc}
We identified point sources in and around each SNR using the SAS task \texttt{edetect\_chain}. Images were extracted from each of the EPIC instruments in the standard energy bands 0.2--0.5~keV, 0.5--1~keV, 1--2~keV, 2--4.5~keV, and 4.5--12~keV, with the same pattern filtering criteria outlined in Section \ref{x-ray-imaging}. False detections due to the extended emission of the SNRs were removed after visual inspection. No point sources were detected within the dimensions of \snra. Two sources were found projected against \snrb: foreground star \object{HD~36877}, the bright source located in the southeast of the elongated remnant emission; and a source with J2000 coordinates of RA~=~05$^{\rm{h}}$27$^{\rm{m}}$58.47$^{\rm{s}}$ and Dec~=~$-71$$^{\rm{d}}$03$^{\rm{m}}$53.7$^{\rm{s}}$, the hard source near the western edge to which we assign the identifier \object{XMMU~J052758.4$-$710353}. Given the hard nature of its emission, XMMU~J052758.4$-$710353 is most likely a background AGN.

\subsection{X-ray spectral analysis}
For the spectral analysis, we made use of the EPIC-pn and EPIC-MOS data. Although less sensitive, the EPIC-MOS spectral resolution is slightly better than the EPIC-pn, and therefore helps to constrain the parameters of the spectral models. Before proceeding with the spectral extraction we generated vignetting-weighted event lists for each EPIC instrument to correct for the effective area variation across our extended source using the SAS task \texttt{evigweight}. Spectra were extracted from elliptical regions encompassing the X-ray extent of the SNRs. Backgrounds were extracted from nearby source and diffuse emission free regions. 

We determined exclusion radii for contaminating point sources (see Section~\ref{src_proc}) using the contour method of the SAS task \texttt{region}. The task calculates the point spread function (PSF) at the source position and normalises for the source brightness. Source counts were then removed down to a PSF threshold of 0.2 times the local background, that is, the point source is excluded down to a level where the surface brightness of the point source is 20\% of the background. This method has the advantage that the point source exclusion regions follow the brightness of the source, so that brighter sources have larger exclusion regions and shapes corresponding approximately to the source PSF. Hence, the number of diffuse counts is maximised by tailoring the point source exclusion regions to the individual sources. We determined exclusion radii of $\sim35\arcsec$ and $\sim18\arcsec$ for HD~36877 and XMMU~J052758.4$-$710353, respectively.

All spectra were rebinned so that each bin contained a minimum of 30 counts to allow the use of the $\chi^{2}$ statistic during spectral fitting. The EPIC-pn and EPIC-MOS source and background spectra from each observation were fitted simultaneously using XSPEC \citep{Arnaud1996} version 12.8.2p with abundance tables set to those of \citet{Wilms2000}, photoelectric absorption cross-sections set to those of \citet{Bal1992}, and atomic data from ATOMDB~3.0.1\footnote{\burl{http://www.atomdb.org/index.php}} with the latest equilibrium and non-equilibrium data.

\subsubsection{X-ray background}
\label{x-ray_background}
A comprehensive description of the X-ray background constituents and spectral modelling can be found in \citet{Maggi2015}. Here we briefly summarise the treatment of the X-ray background in the cases of \snra\ and \snrb.

The X-ray background consists of the astrophysical X-ray background (AXB) and particle induced background. The AXB typically comprises four or fewer components \citep{Snowden2008,Kuntz2010}, namely the unabsorbed thermal emission from the Local Hot Bubble, absorbed cool and hot thermal emission from the Galactic halo, and an absorbed power law representing unresolved background active galactic nuclei (AGN). The spectral properties of the background AGN component were fixed to the well known values of $\Gamma \sim 1.46$ and a normalisation equivalent to 10.5 photons~cm$^{-2}$~s$^{-1}$~sr$^{-1}$ at 1~keV  \citep{Chen1997}. The foreground absorbing material comprises both Galactic and LMC components. The foreground Galactic absorption component was fixed at $7.9\times10^{20}$~cm$^{-2}$ for \snra\ and $6.9\times10^{20}$~cm$^{-2}$ for \snrb, based on the \citet{Dickey1990} \ion{H}{i} maps, determined using the HEASARC $N_{\rm{H}}$ Tool\footnote{\burl{http://heasarc.gsfc.nasa.gov/cgi-bin/Tools/w3nh/w3nh.pl}}. The foreground LMC absorption component, with abundances set to those of the LMC, was allowed to vary. 


The particle-induced background of the EPIC consists of the quiescent particle background (QPB), instrumental fluorescence lines, electronic read-out noise, and residual soft proton (SP) contamination. To determine the contribution of these components we made use of vignetting corrected FWC data. We extracted FWC spectra from the same detector regions as the observational source and background spectra. The EPIC-pn and EPIC-MOS FWC spectra were fitted with the empirical models developed by \citet{SturmPhD} and \citet{Maggi2015}, respectively. Since these spectral components are not subject to the instrumental response, we used a diagonal response in XSPEC. The resulting best-fit model was included and frozen in the fits to the observational spectra, with only the widths and normalisations of the strongest fluorescence lines allowed to vary\footnote{Following the method outlined in the \xmm\ Extended Source Analysis Software Cookbook available at \burl{http://heasarc.gsfc.nasa.gov/docs/xmm/xmmhp_xmmesas.html}}. We also included a multiplicative constant to normalise the continuum to the observational spectra using the high energy tail ($E>5$~keV) where the QPB component dominates. The residual SP contamination was fitted by a power law not convolved with the instrumental response \citep{Kuntz2008}. The residual SP component was only required in the spectral analysis of \snra.

\subsubsection{Spectral fitting: \snra}
\noindent The X-ray emission of \snra\ consists of a centrally bright region peaking in the 0.7--1.1~keV range, surrounded by a softer shell (see Fig.~\ref{1139_4panel}-top left). A similar morphology has been observed in other LMC remnants and is discussed in detail in Section~\ref{1139-mwm}. The centrally peaked 0.7--1.1~keV emission suggests that the Fe L-shell complex dominates at these energies. Inspection of the spectrum confirmed this to be the case with an obvious Fe~L-shell `bump' present. The softer outer emission is most likely due to a swept-up and shock-heated ISM shell, typical of evolved SNRs. The separation of core and shell regions is not straightforward. As seen in Fig.~\ref{1139_4panel}-top left, the 0.7--1.1~keV emission is not confined to a central well-defined core, but rather is somewhat smeared out from the core in both northernly and southernly directions. It is not clear if this emission is due to Fe or is simply caused by enhancements in the shell emission. The latter may be the case in the northern extension from the core, which is correlated with bright soft emission. However, in the southern extension the 0.7--1.1~keV emission is anti-correlated with the soft shell. Since we cannot define clear core and shell regions, we decided to fit the spectrum of the entire SNR, incorporating both shell and core components. We note here that we also tried to incorporate additional pure heavy element plasmas such as O to account for ejecta emission apart from Fe. However, in fits to the spectrum of the entire remnant and the test spectra extracted from the brightest core regions, additional ejecta components were not required with only upper limits determined. Therefore, we consider only the Fe ejecta component in the forthcoming analysis.

To account for the shell emission, we included a thermal plasma model in the spectral fits absorbed by foreground Galactic and LMC material. We made use of a non-equilibrium ionisation (NEI) model in XSPEC, appropriate for SNRs, namely the plane-parallel \texttt{vpshock} model \citep{Borkowski2001}. The \texttt{vpshock} model features a linear distribution of ionisation ages behind the shock.
Given the large size of the \snra\ (see Section~\ref{1139-mwm}), we assumed that the remnant is in the Sedov phase and the emission from the soft X-ray shell is dominated by swept-up ISM. Therefore, we also fit the spectrum with the Sedov dynamical model, implemented as \texttt{vsedov} in XSPEC \citep{Borkowski2001}.



\noindent To account for the central Fe L-shell emission we used a pure Fe thermal plasma model, achieved by setting all abundances apart from Fe to 0. We performed trial fits with CIE models and NEI models. We note that for the CIE fits we did not use the standard \vapec\ model. The reason for this is because it not possible to set the H abundance to 0 in the \vapec\ model. In past versions of XSPEC, this was not a problem as the abundance of Fe could be set sufficiently high to render any contribution of H insignificant. However, in the most recent versions of XSPEC, the abundance of elements in the \vapec\ model has a much lower maximum limit and the effect of H on the spectral fits can no longer be ignored. To circumvent this issue, we used the NEI model \vnei\ and froze the ionisation parameter to $1\times10^{12}$~s~cm$^{-3}$, consistent with a plasma in CIE \citep{Smith2010}. The \vnei\ model does allow the value abundance of H to be set to 0, therefore providing a pure Fe plasma model. The CIE fits produced a good fit to the core Fe emission. Similarly, the ionisation parameters in the NEI fits were consistently $\gtrsim10^{12}$~s~cm$^{-3}$, indicative of a CIE plasma. Therefore, we adopted the CIE model to account for the Fe L-shell emission.

\begin{figure*}
\begin{center}
\resizebox{\hsize}{!}{\includegraphics[trim= 0cm 0cm 0cm 0cm, clip=true, angle=0]{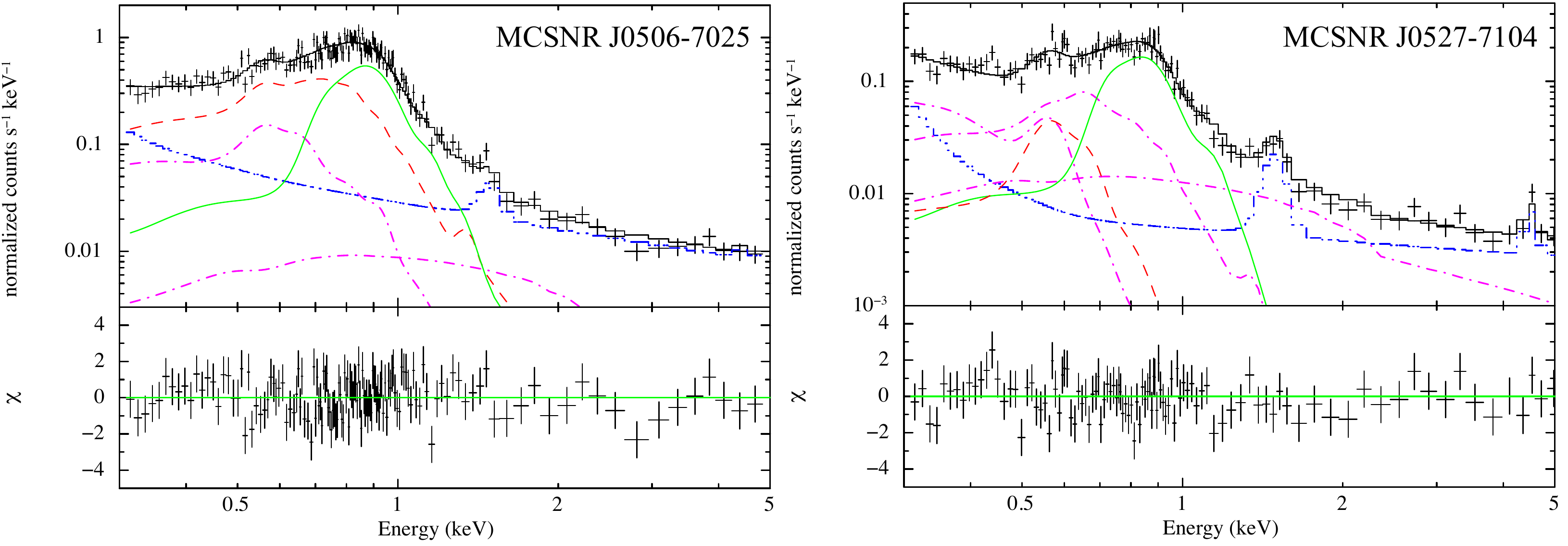}}
\caption{{\bf Left:} Best-fit \vsedov+\vnei\ model to the X-ray spectrum of \snra. The green solid line represents the pure Fe component, the dashed red line shows shell component, the magenta dash-dot lines mark the AXB components, and the blue dash-dot-dot-dot line shows the combined contributions of the QPB, instrumental fluorescence lines, and electronic noise. {\bf Right:} Best-fit \vnei+\vnei\ model to the X-ray spectrum of \snrb. Same as left except the dashed red line represents the pure O component. In both cases, only the EPIC-pn spectrum is shown for clarity. Best-fit parameters for \snra\ and \snrb\ are given in Tables~\ref{1139_tab} and \ref{1234_tab}, respectively. }
\label{specs}
\end{center}
\end{figure*}

Both the \vpshock+\vnei\ and \vsedov+\vnei\ models provided good fits to the \snra\ spectrum with reduced $\chi^{2}$ ($\chi^{2}_{\nu}$) of 1.16 and 1.11 respectively. The best-fit parameters for both models are given in Table~\ref{1139_tab}, with the best-fit \vsedov+\vnei\ spectrum shown in Fig.~\ref{specs}-left. These fit results are discussed in Section~\ref{1139-xray-analysis}.
\\

\subsubsection{Spectral fitting: \snrb}
\noindent The X-ray morphology of \snrb\ is extremely unusual (see Section~\ref{1234-mwm} for a thorough discussion) with the emission almost exclusively confined to the 0.7--1.1~keV band (see Fig.~\ref{1234_4panel}-top left), suggesting that the spectrum could be dominated by emission lines from the Fe~L-shell complex, which was confirmed upon inspection of the spectrum. There is no evidence of soft X-ray emission filling out the optical shell of \snrb\ (see Fig.~\ref{1234_4panel}-bottom left). We therefore proceeded by applying a single thermal component consisting of pure Fe. Initial trial fits with NEI models resulted in ionisation parameters $\gtrsim10^{12}$~s~cm$^{-3}$, consistent with a plasma in CIE. For this reason we adopted a pure Fe \vnei\ model with the ionisation parameter fixed at $1\times10^{12}$~s~cm$^{-3}$. This yielded an acceptable fit with $\chi^{2}_{\nu}$~=~1.20, however, residuals remained in the 0.5--0.7~keV range, suggesting the presence of O. Therefore, we added a second pure O \vnei\ component to the model, with $\tau$ again fixed to $1\times10^{12}$~s~cm$^{-3}$, resulting in an improved fit with $\chi^{2}_{\nu}$~=~1.11, shown in Fig.~\ref{specs}-right with fit results given in Table~\ref{1234_tab}. The physical interpretation of the results is discussed in Section~\ref{1234-disc}.

\begin{table}
\caption{Spectral fit results for \snra. See text for description of the models.}
\begin{center}
\label{1139_tab}
\begin{tabular}{llr}
\hline
Component & Parameter & Value\\
\hline
\hline
\multicolumn{3}{c}{\vpshock\ (shell) + \vnei\ (Fe core)}\\
\hline
\multicolumn{3}{c}{ }\\
\phabs & $N_{\rm{H,Gal}}$ ($10^{22}$ cm$^{-2}$) &   0.08\tablefootmark{a}  \\
\vphabs\tablefootmark{b}  & $N_{\rm{H,LMC}}$($10^{22}$ cm$^{-2}$) &   $<0.01$  \\
\multicolumn{3}{c}{ }\\
\vpshock\tablefootmark{b}\ & $kT$ & 0.48 (0.30--0.74)\\
 & $\tau_{u}$ ($10^{11}$~s~cm$^{-3}$) & 0.9 (0.4--3.0) \\
 & $EM$ ($10^{57}$ cm$^{-3}$) & 1.7 (0.92--3.65) \\
\multicolumn{3}{c}{ }\\
\vnei & $kT_{\rm{Fe}}$ & 0.78 (0.76--0.79) \\
  & $\tau$ ($10^{12}$~s~cm$^{-3}$) & 1.0 (fixed) \\
 & $n_{e}n_{\rm{Fe}}V$ ($10^{53}$ cm$^{-3}$) & 1.15 (1.12--1.24) \\ 
\multicolumn{3}{c}{ }\\
 & $F_{X, total}$\tablefootmark{c} ($10^{-13}$ erg~s$^{-1}$~cm$^{-2}$) & 4.4 \\
 & $L_{X, shell}$\tablefootmark{d} ($10^{34}$ erg~s$^{-1}$) & 8.4 \\ 
 & $L_{X, Fe}$\tablefootmark{d} ($10^{34}$ erg~s$^{-1}$) & 9.5 \\
 \multicolumn{3}{c}{ }\\
Fit statistic & $\chi^{2}_{\nu}$ & 1.16 (472 d.o.f) \\
\multicolumn{3}{c}{ }\\
\hline
\hline
\multicolumn{3}{c}{\vsedov\ (shell) + \vnei\ (Fe core)}\\
\hline
\multicolumn{3}{c}{ }\\
\phabs & $N_{\rm{H,Gal}}$ ($10^{22}$ cm$^{-2}$) &   0.08\tablefootmark{a}  \\
\vphabs\tablefootmark{b} & $N_{\rm{H,LMC}}$ ($10^{22}$ cm$^{-2}$) &   $<0.07$  \\
\multicolumn{3}{c}{ }\\
\vsedov\tablefootmark{b}\ & $kT$ & 0.18 (0.15--0.37) \\
 & $\tau_{u}$ ($10^{11}$~s~cm$^{-3}$) & 4.89 ($>2.49$) \\
 & $EM$ ($10^{58}$ cm$^{-3}$) & 1.07 (0.28--1.49) \\
\multicolumn{3}{c}{ }\\
\vnei & $kT_{\rm{Fe}}$ & 0.80 (0.77--0.82) \\
  & $\tau$ ($10^{12}$~s~cm$^{-3}$) & 1.0 (fixed) \\
 & $n_{e}n_{\rm{Fe}}V$ ($10^{53}$ cm$^{-3}$) & 1.09 (0.97--1.20) \\ 
\multicolumn{3}{c}{ }\\
 & $F_{X, total}$\tablefootmark{c} ($10^{-13}$ erg~s$^{-1}$~cm$^{-2}$) & 4.5 \\ 
 & $L_{X, shell}$\tablefootmark{d} ($10^{35}$ erg~s$^{-1}$) & 1.8 \\ 
 & $L_{X, Fe}$\tablefootmark{d} ($10^{34}$ erg~s$^{-1}$) & 8.8 \\
 \multicolumn{3}{c}{ }\\
Fit statistic & $\chi^{2}_{\nu}$ & 1.11 (472 d.o.f) \\
\multicolumn{3}{c}{ }\\
\hline
\end{tabular}
\tablefoot{The numbers in parentheses are the 90\% confidence intervals.
\tablefoottext{a}{Fixed to the Galactic column density from the \citet{Dickey1990} \ion{H}{i} maps.}
\tablefoottext{b}{Absorption and thermal component abundances fixed to those of the LMC.}
\tablefoottext{c}{0.3-10~keV absorbed X-ray flux.}
\tablefoottext{d}{0.3-10~keV de-absorbed X-ray luminosity, adopting a distance of 50~kpc to the LMC.}
}
\end{center}
\end{table}%

\begin{table}
\caption{Spectral fit results for \snrb. See text for description of the model.}
\begin{center}
\label{1234_tab}
\begin{tabular}{llr}
\hline
Component & Parameter & Value\\
\hline
\hline
\multicolumn{3}{c}{\vnei\ + \vnei}\\
\hline
\multicolumn{3}{c}{ }\\
\phabs & $N_{\rm{H,Gal}}$ ($10^{22}$ cm$^{-2}$) &   0.07\tablefootmark{a}  \\
\vphabs\tablefootmark{b}  & $N_{\rm{H,LMC}}$ ($10^{22}$ cm$^{-2}$) &   $<0.05$ \\
\multicolumn{3}{c}{ }\\
\vnei\ $(Fe)$ & $kT_{\rm{Fe}}$ & 0.71 (0.68--0.73) \\
  & $\tau$ ($10^{12}$~s~cm$^{-3}$) & 1.0 (fixed) \\
 & $n_{e}n_{\rm{Fe}}V$ ($10^{52}$ cm$^{-3}$) & 2.7 (2.5--2.9) \\ 
\multicolumn{3}{c}{ }\\
 \vnei\ $(O)$ & $kT_{\rm{O}}$ & 0.18 ($0.14-0.23$) \\
  & $\tau$ ($10^{12}$~s~cm$^{-3}$) & 1.0 (fixed) \\
 & $n_{e}n_{\rm{O}}V$ ($10^{53}$ cm$^{-3}$) & 2.2 (1.6--3.8) \\ 
\multicolumn{3}{c}{ }\\
 & $F_{X}$\tablefootmark{c} ($10^{-14}$ erg~s$^{-1}$~cm$^{-2}$) & 7.3 \\
 & $L_{X, Fe}$\tablefootmark{d} ($10^{34}$ erg~s$^{-1}$) & 2.2 \\ 
 & $L_{X, O}$\tablefootmark{d} ($10^{34}$ erg~s$^{-1}$) & 0.5 \\
 
\multicolumn{3}{c}{ }\\
Fit statistic & $\chi^{2}_{\nu}$ & 1.11 (475 d.o.f.) \\
\multicolumn{3}{c}{ }\\
\hline
\end{tabular}
\tablefoot{The numbers in parentheses are the 90\% confidence intervals.
\tablefoottext{a}{Fixed to the Galactic column density from the \citet{Dickey1990} \ion{H}{i} maps.}
\tablefoottext{b}{Absorption component abundances fixed to those of the LMC.}
\tablefoottext{c}{0.3-10~keV absorbed X-ray flux.}
\tablefoottext{d}{0.3-10~keV de-absorbed X-ray luminosity, adopting a distance of 50~kpc to the LMC.}
}
\end{center}
\end{table}%



\section{Results}
\label{results}
\subsection{\snra}
\subsubsection{Multi-wavelength morphology}
\label{1139-mwm}
The X-ray emission from \snra\ is characterised by a bright core in the 0.7--1.1~keV range, surrounded by a soft shell evident in the 0.3--0.7~keV band (Fig.~\ref{1139_4panel}-top left). The X-ray spectral analysis has shown that the enhanced 0.7--1.1~keV emission in the core is due to an Fe rich gas, most likely reverse shock heated Fe ejecta. Similarly, the soft shell is consistent with swept-up and shock-heated ISM, typical of an evolved SNR. This Fe core/soft shell morphology has been observed in several other evolved LMC SNRs, including DEM~L238 and DEM~L249 \citep{Borkowski2006} and MCSNR~J0508--6902 \citep{Bozzetto2014}. The 0.7--1.1~keV emission is not confined to a clear, well-defined core, with emission spreading in both northernly and southernly directions. This may be due to ejecta from the Fe core being smeared out across the remnant, or simply mark an enhancement of the shell emission. While the northern 0.7--1.1~keV emission is largely correlated with the softer shell emission, the southern extension is not, suggesting that the Fe gas is not confined to a well-defined core. 

\begin{figure}[!t]
\begin{center}
\resizebox{\hsize}{!}{\includegraphics[trim= 0.2cm 0.7cm 2cm 2cm, clip=true, angle=0]{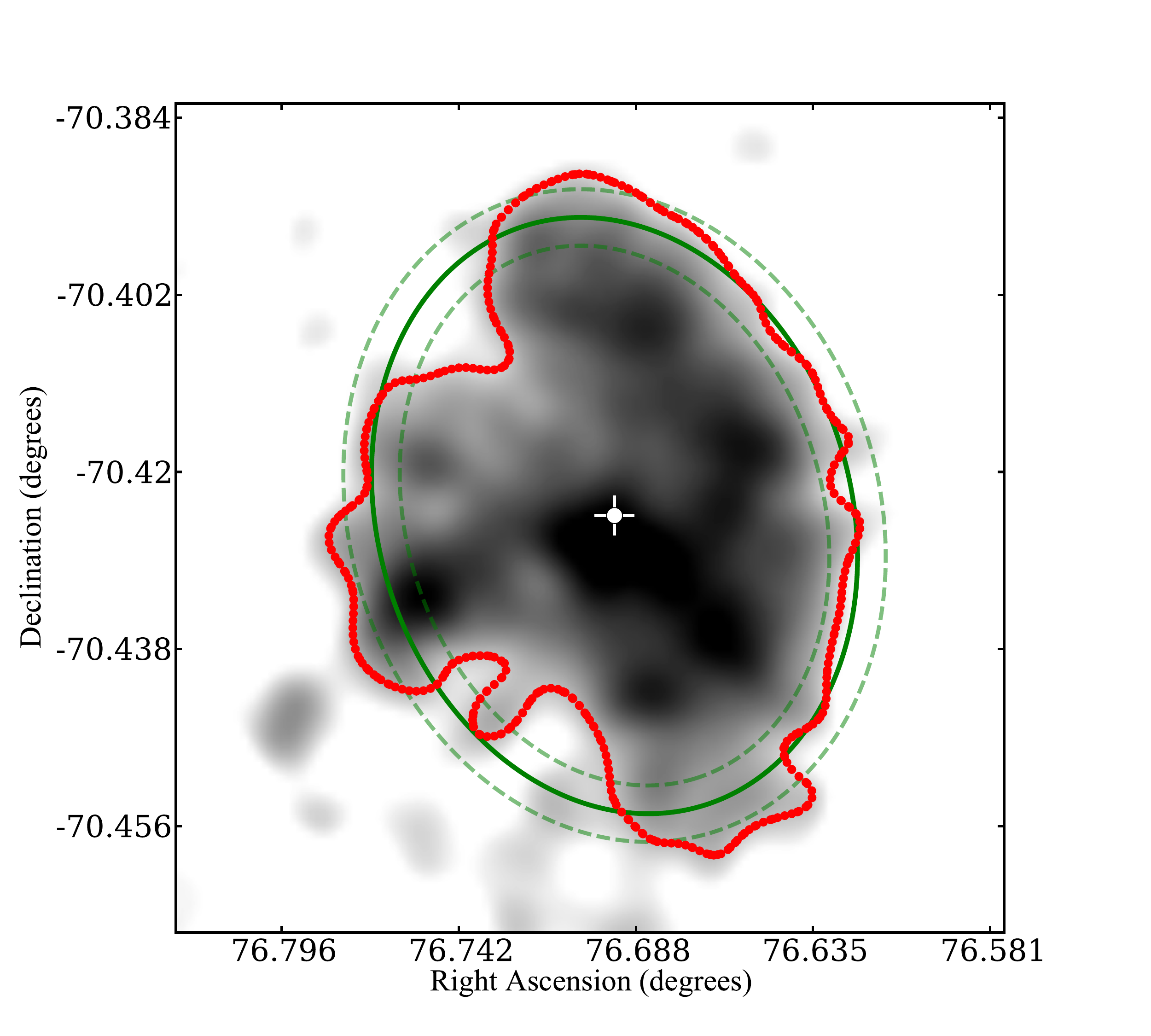}}
\caption{Combined 0.3--0.7~keV EPIC image of \snra. The red points delineate the contour level corresponding to 3$\sigma$ above the average background surface brightness. The green solid line shows the best-fit ellipse to the contour, with the dashed lines indicating the 1$\sigma$ error on the fit. The white plus-sign marks the best-fit centre of the SNR.}
\label{1139_size-fit}
\end{center}
\end{figure}

\begin{figure*}
\begin{center}
\resizebox{\hsize}{!}{\includegraphics[trim= 0cm 0cm 0cm 0cm, clip=true, angle=0]{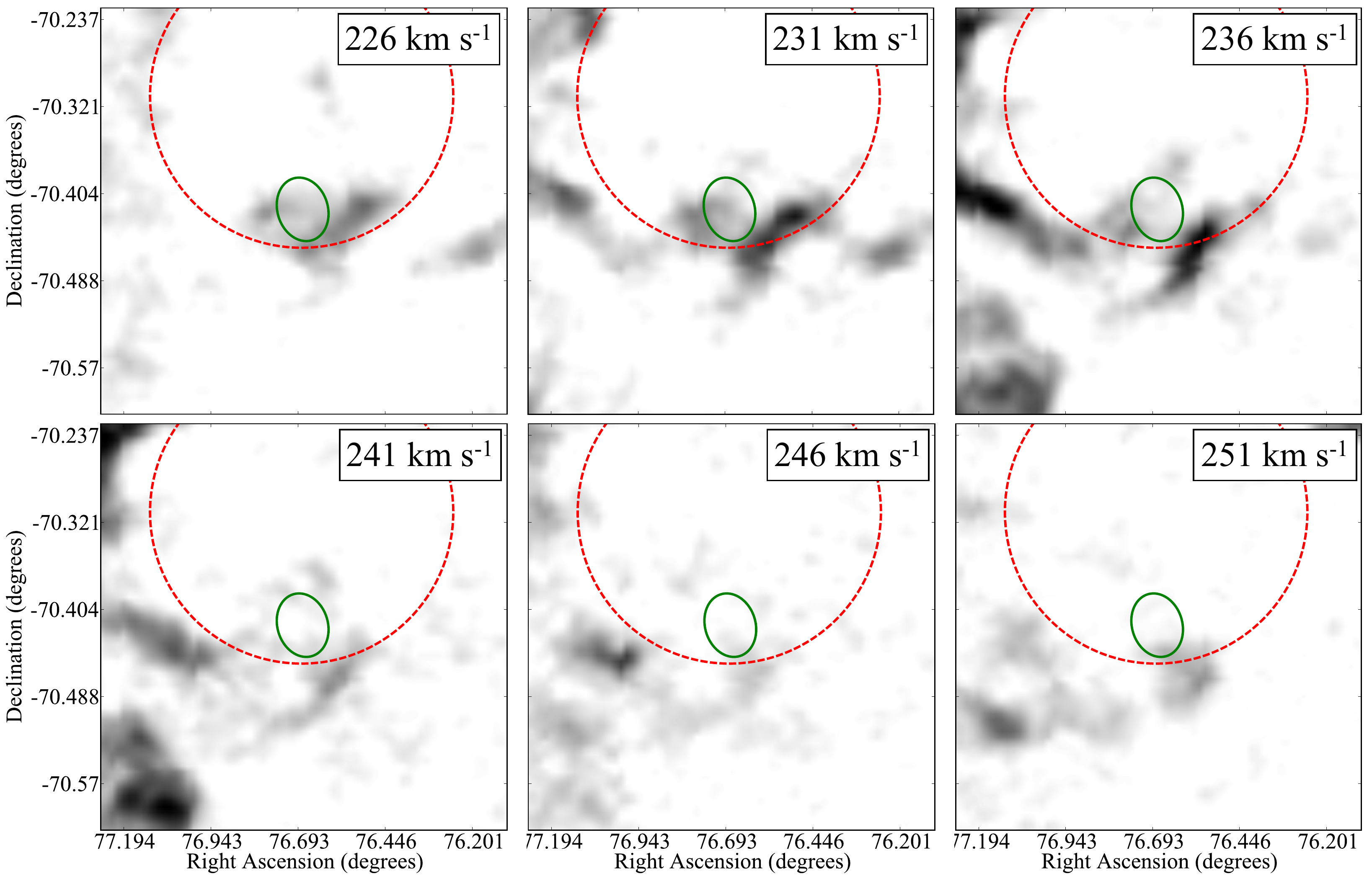}}
\caption{\ion{H}{i} channel maps of the \snra\ region. The dimensions of the SNR are shown by the green ellipse in each panel. The red dashed line mark the approximate position of the giant \ion{H}{i} shell \object{GS~30} \citep{Kim1999}. The heliocentric velocities are also indicated.}
\label{1139_HI}
\end{center}
\end{figure*}

The shell emission is elliptical, more extended in the north-south direction than east-west. There are some brightness enhancements, notably in a knot at the very east and along the western front. These regions are well correlated with enhanced radio emission (Fig.~\ref{radioemission}), and regions of highest polarisation (Fig.~\ref{6cmpol}). Well-defined shell-like \sii\ enhancements (see Fig.~\ref{1139_4panel}-top right) corresponding to regions where \ratio$>0.67$, are detected ahead of the correlated X-ray and radio emission along the western front. These morphologies are consistent with the blast wave having encountered some denser ambient medium in these regions. To investigate this we made use of data from the SAGE survey of the LMC (see Section~\ref{ir}). The 24~$\mu$m image of \snra\ is shown in Fig.~\ref{1139_4panel}-bottom right, and reveals enhanced dust emission ahead of the eastern, western, and southern fronts. We investigated this further using data from the \ion{H}{i} survey of the LMC (see Section~\ref{hi}). We extracted channel maps centred on \snra, shown in Fig.~\ref{1139_HI}. The SNR is located at the GS~30 \ion{H}{i} shell \citep{Kim1999}, with distribution of \ion{H}{i} in the 226~km~s$^{-1}$, 231~km~s$^{-1}$, 236~km~s$^{-1}$ channel maps consistent with the expectation of ambient density enhancements to the east and west of \snra. Along the western front, the magnetic field vectors are oriented tangential to the shock front, typical of older remnants where there is compression of the interstellar ambient field. The field vectors in the knot to the east show a radial orientation. We also note that faint radio emission was found to extend $\sim1\arcmin-2\arcmin$ beyond the observed X-ray shell to the south (see Figs.~\ref{radioemission} and \ref{6cmpol}). It is unclear if this is associated with \snra\ or is unrelated background emission, possibly due to the nearby \ion{H}{ii} region. The radio spectral index map shown in Fig.~\ref{spchist}-left indicates a flatter spectrum in this feature, more consistent with a thermal origin. For this reason, and given the absence of correlated X-ray and characteristic optical emission, it is likely that the extended radio feature is unrelated to the SNR.

To estimate the size of the X-ray remnant we determined the average background surface brightness and corresponding standard deviation ($\sigma$) in the 0.3--0.7~keV band. We then defined the edge of the SNR as regions where the extended emission surface brightness rises to 3$\sigma$ above the average background and fitted an ellipse to this contour. The error on the fit was determined by quantifying the standard deviation of points on the contour from the best-fit ellipse. We determined a best-fit ellipse centred on the J2000 coordinates RA~=~05$^{\rm{h}}$06$^{\rm{m}}$47.30$^{\rm{s}}$ and Dec~=~$-70$$^{\rm{d}}$25$^{\rm{m}}$29.7$^{\rm{s}}$, of size $2\farcm89~(\pm0\farcm35)\times3\farcm71~(\pm0\farcm35)$, corresponding to $42.1~(\pm5.0)~\rm{pc}\times54.0~(\pm5.0)$~pc at the LMC distance, with the major axis rotated $\sim17^{\circ}$ east of north. The best-fit dimensions and error are shown in Fig.~\ref{1139_size-fit}.

\subsubsection{X-ray emission}
\label{1139-xray-analysis}
The X-ray spectrum of \snra\ was found to be well described by either the \vpshock+\vnei\ or \vsedov+\vnei\ model with marginal difference in the fit-statisitic ($\chi^{2}_{\nu}=1.16$ and $\chi^{2}_{\nu}=1.11$, respectively), as shown in Table~\ref{1139_tab}. In each case, the best-fit absorption due to foreground material in the LMC was negligible, with only upper limits determined. As argued in \citet{Maggi2015}, the negligible foreground absorption can be interpreted as the SNR being located on the near side of the LMC. Therefore, it is possible that the apparent correlation of \snra\ with GS~30, shown in Fig.~\ref{1139_HI}, is simply a chance alignment. Moreover, we argue in the forthcoming sections for a Type~Ia origin for \snra\ and, therefore, the association of the remnant with an \ion{H}{i} giant shell is not expected. However, given the correlation between the bright regions of X-ray material with denser material in the \ion{H}{i} maps (see Section~\ref{1139-mwm}, it is likely that there is at least some interaction of the remnant and GS~30.

We now discuss the physical interpretation of the shell and ejecta emission components.\\

\noindent {\it Shell emission:} \\
\noindent The best-fit plasma temperatures for the shell components of each model agree within the 90\% confidence intervals, and are consistent with other evolved SNRs in the LMC \citep[see][for example]{Maggi2014,Kavanagh2015b}. The shell ionisation parameters inferred from the fits are also compatible, though only a lower limit could be determined for the \vsedov\ component. 

Using our fit results we can estimate physical parameters of \snra\ using the Sedov dynamical model \citep[see][for examples]{Sasaki2004,Kavanagh2015b}. The X-ray shell of the remnant is delineated by an ellipse (see Section~\ref{1139-mwm}), with semi-major and semi-minor axes of 27.0~pc and 21.0~pc, respectively.  Assuming these are the first and second semi-principal axes of an ellipsoid describing \snra, and that the third semi-principal axis is in the range 27.0--21.0 pc, we determined the volume ($V$) limits for the remnant and their corresponding effective radii ($R_{\rm{eff}}$) to be $(1.5-1.9)\times10^{60}$ cm$^{3}$ and 22.8--24.8~pc, respectively.\\

\par The best fit shell X-ray temperature (see Table~\ref{1139_tab}) corresponds to a shock velocity

\begin{equation}
v_{s}=\left(\frac{16kT_{s}}{3\mu}\right)^{0.5},
\end{equation}

\noindent where $kT_{s}$ is the postshock temperature and $\mu$ is the mean mass per free particle. For a fully ionised plasma with LMC abundances, $\mu=0.61\rm{m_{p}}$, resulting in a shock velocity of $v_{s}~=~388~(354-557)$~km~s$^{-1}$. The age of the remnant can now be determined from the self-similarity solution:

\begin{equation}
v_{s}=\frac{2R}{5t},
\end{equation}

\noindent where $R=R_{\rm{eff}}$ and $t$ is the age of the remnant. This gives an age range of $16-28$~kyr. The pre-shock H density ($n_{\rm{H},0}$) in front of the blast wave can be determined from the emission measure ($EM$, see Table \ref{1139_tab}). Evaluating the emission integral over the Sedov solution using the approximation for the radial density distribution of \citet{Kahn1975} gives

\begin{equation}
\int n_{e}n_{\rm{H}}dV=EM=2.07\left(\frac{n_{e}}{n_{\rm{H}}}\right)n_{\rm{H},0}^{2}V,
\end{equation}

\noindent where $n_{\rm{e}}$ and $n_{\rm{H}}$ are electron and hydrogen densities, respectively, V is the volume \citep[e.g.,][]{Hamilton1983}. Taking $n_{e}/n_{\rm{H}}=1.21$ and the determined range of volumes, this equation yields $n_{\rm{H},0}=(2.4-6.4)\times10^{-2}$~cm$^{-3}$. Since the pre-shock density of nuclei is given as $n_{0}\sim1.1 n_{\rm{H},0}$, it follows that  $n_{0} = (2.6-7.0)\times10^{-2}$~cm$^{-3}$. Hence, the SNR is expanding into a fairly rarified environment. The initial explosion kinetic energy ($E_{0}$) can be determined from the equation:

\begin{equation}
R=\left(\frac{2.02E_{0}t^{2}}{\mu_{n}n_{0}}\right)^{1/5},
\end{equation}

\noindent where $\mu_{n}$ is the mean mass per nucleus ($=1.4m_{p}$). This results in an initial explosion energy in the range $(0.07-0.84)\times10^{51}$~erg. This is lower than the canonical $10^{51}$~erg but consistent with several Type Ia SNRs in the LMC \citep[e.g.,][]{Hendrick2003,Bozzetto2014}. The swept-up mass contained in the shell is given simply by $M=V\mu_{n}n_{0}$, which is evaluated to ($46-155$)~M$_{\odot}$. This large amount of swept-up material supports our original assumption that \snra\ is into the Sedov phase of its evolution. All the derived parameters of \snra\ are summarised in Table \ref{1139_prop}.

\begin{table}[t]
\caption{Physical properties of \snra\ derived from the Sedov model}
\begin{center}
\begin{normalsize}
\label{1139_prop}
\begin{tabular}{ccccc}
\hline
$n_{\rm{0}}$ & $v_{s}$ & $t$ & $M$ & $E_{0}$ \\
 ($10^{-2}$~cm$^{-3}$) & km~s$^{-1}$ & (kyr) & (M$_{\sun}$) & ($10^{51}$ erg) \\
\hline
\hline
2.6 -- 7.0 & 354 -- 557 & 16 -- 28  & 46 -- 155 & 0.07 -- 0.84 \\
\hline
\end{tabular}
\end{normalsize}
\end{center}
\end{table}%

Using the determined values of $t$ and $n_{0}$ we estimated the expected value of the ionisation parameter ($\tau$) in the \vsedov\ component of the spectral fit. Given that $\tau$ is the product of the electron density immediately behind the shock front and the remnant age \citep{Borkowski2001}, it can be written $\tau = n_{e}t \approx 4.8n_{\rm{H},0}t$. We determined the expected $\tau$ to be in the range $(0.60-2.52)\times10^{11}$~cm$^{-3}$~s, which is in agreement with the $\tau$ determined in the spectral fits (see Table~\ref{1139_tab}), though the fitted ionisation parameter is poorly constrained.\\

\noindent{\it Ejecta emission:} \\
\noindent The interior emission of \snra\ was modelled with a thermal plasma model consisting of pure Fe. We found that a CIE plasma with best-fit of $kT_{\rm{Fe}} = 0.78~(0.76-0.79)$~keV in the \vpshock+\vnei\ model and $kT_{\rm{Fe}} = 0.80~(0.77-0.82)$~keV in the \vsedov+\vnei\ model provided a good fit to the spectrum (see Table~\ref{1139_tab}). The plasma temperatures and emission measures for both models are in agreement within the 90\% confidence intervals. We proceed adopting the Fe emission parameters of the \vsedov+\vnei\ model. The mass of Fe in the core can be determined from the emission measure of the gas and the volume. This method is described in detail in \citet{Kosenko2010,Bozzetto2014,Maggi2014}, with the Fe mass given by the equation

\begin{equation}
\label{Fe-calc}
M_{\rm{Fe}}  = (V_{\rm{Fe}} EM_{\rm{Fe}})^{0.5} (n_{\rm{e}}/n_{\rm{Fe}})^{-0.5} m_{\rm{U}} A_{\rm{Fe}},
\end{equation}

\noindent where $V_{\rm{Fe}}$ is the volume occupied by the Fe, $EM_{\rm{Fe}}=n_{e}n_{Fe}V$ is the emission measure of the Fe gas, $n_{\rm{e}}/n_{\rm{Fe}}$ is the electron to Fe-ion ratio, $m_{\rm{U}}$ is the atomic mass unit, and $A_{\rm{Fe}}$ is the atomic mass of Fe. $EM_{\rm{Fe}}$ is obtained from the normalisation of the Fe component and is listed in Table \ref{1139_tab}. The $n_{\rm{Fe}}/n_{\rm{H}}$ value is calculated from the Fe abundance parameter of the model. 

\begin{figure}[!t]
\begin{center}
\resizebox{\hsize}{!}{\includegraphics[trim= 0cm 0cm 0cm 0cm, clip=true, angle=0]{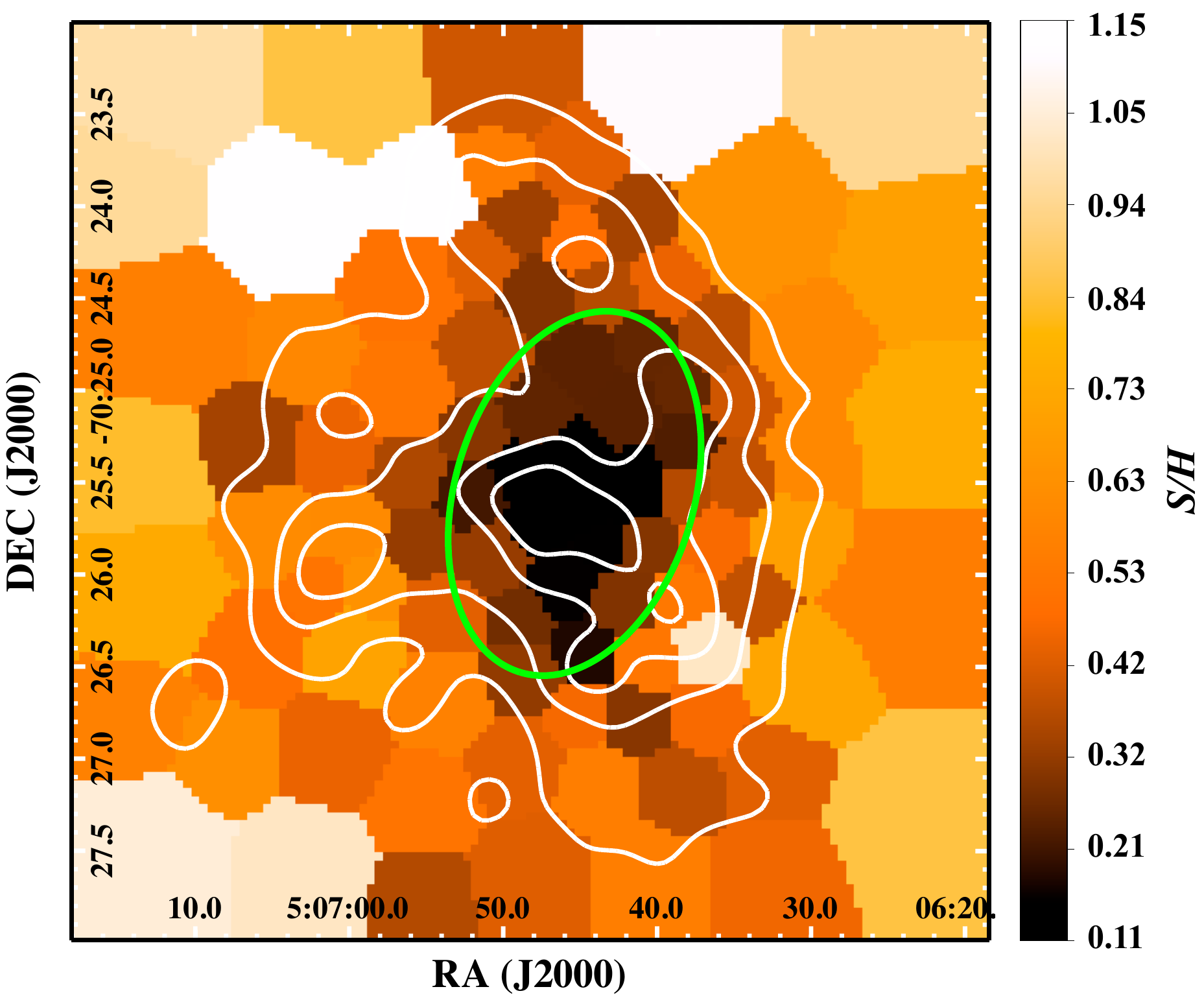}}
\caption{WVT X-ray colour map of \snra. The tessellates are defined so that each has a S/N of 5. Lower values of $S/H$ indicate higher contributions from Fe L-shell emission. The white contours show the 0.3--0.7~keV counters from Fig.~\ref{1139_4panel}-top right and the green ellipse marks the Fe L-shell emission regions, set to $S/H < 0.3$ (see text).}
\label{wvt}
\end{center}
\end{figure}

The determination of $V_{\rm{Fe}}$ presents a problem. As discussed in Section~\ref{1139-mwm}, the emission in the 0.7--1.1~keV range is not confined to a central well-defined core, with extensions evident to the north and south. It is unclear if these extensions are due to the Fe-rich ejecta or to enhancements in the shell emission. Ideally, we could perform a spatially resolved spectral analysis from various regions around the remnant to determine where an Fe L-shell bump is present. Unfortunately, because of the count statistics, this is not possible with the current data set. However, we can get an estimate of the distribution of Fe by comparing count rates ($N$) in specific energy bands across the remnant and creating an X-ray colour map. To do this we made use of the Weighted Voronoi Tessellation (WVT) binning algorithm by \citet{Diehl2006}\footnote{See also \burl{http://www.phy.ohiou.edu/~diehl/WVT/index.html\#home}.}, which is a generalisation of the \citet{Capp2003} Voronoi binning algorithm. This is an adaptive image binning technique that uses weighted Voronoi tessellations to bin the image to a user defined signal-to-noise (S/N) per bin while keeping the bin sizes as compact as possible. The WVT binning tools provide an X-ray colour map facility which compares the counts in two energy bands to gain a picture of the variation in X-ray colour over the remnant. We defined the two energy bands, a soft band $S = 0.3-0.7$~keV (where the shell dominates) and a hard band $H = 0.7-1.0$~keV (where the Fe L-shell bump peaks), and the colour $S/H$ is then given by $N_{0.3-0.7~\rm{keV}}/N_{0.7-1.0~\rm{keV}}$. We set a target S/N of 5 so that the bins give a useful estimate of $S/H$ in each bin while producing small enough bins to get a picture of the Fe distribution. The resulting WVT X-ray colour map is shown in Fig.~\ref{wvt}. It is already clear from the image that Fe dominates the core, as expected. However, we must decide where to apply the cut to the Fe dominating regions. To do this we performed spectral simulations assuming that the best-fit core and shell component parameters are constant over the remnant (see Table~\ref{1139_tab}), allowing only the relative normalisation between the shell and core components to vary. These simulations suggested that the pure Fe component becomes significant when $S/H$ drops below $\sim0.35$. Given the relatively low S/N per bin, we set a conservative upper limit at $S/H < 0.3$. The majority of tessellates with $S/H < 0.3$ are aggregated in the centre of the SNR. We traced an ellipse around these tessellates by eye, and assume that this represents the projected volume of the Fe ejecta (green ellipse in Fig.~\ref{wvt}). The resulting dimensions of the central region are $\sim19.0~\rm{pc}\times29.6~\rm{pc}$. We assume an ellipsoidal morphology with a third semi-axis equal to the mean of the semi-major and semi-minor axes ($12.2$~pc). If the actual morphology is oblate or prolate, the volume would be $\sim22$\% higher or lower, respectively. 

As discussed in \citet{Hughes2003}, there are two limiting cases for the value of the $n_{\rm{e}}/n_{\rm{Fe}}$ ratio, depending on the level of admixture of H in the ejecta. If there is no H in the ejecta, then the number of free electrons per Fe ion only depends on the average ionisation state of the Fe. For a plasma in CIE at $kT= 0.80$~keV (log $T~\sim~6.95$) the average ionisation is 18.3 for Fe \citep{Shull1982}. Then, from Equation \ref{Fe-calc}, the Fe mass is in the range 1.6--1.7 $M_{\sun}$. Alternatively, one can assume that a similar mass of H is mixed into the Fe-rich ejecta. Therefore, the number density of Fe over H is 1/56, and the average number of electrons per Fe ion is 74.3. It follows from Eq.~\ref{Fe-calc} that the Fe mass is in the range 0.8--0.9 $M_{\sun}$ if there is a comparable mass of H to Fe in the ejecta, slightly higher than expected. However, the ejecta may not occupy the entire volume of the projected morphology which would be expected if the ejecta were clumpy. In this case, the emitting volume must be modified by a filling factor of $\sim0.4$ \citep[see][and references therein]{Kosenko2010}, in which case the Fe content reduces by a factor of $\sqrt{0.4}\approx0.63$ giving 1.0--1.1 $M_{\sun}$ and 0.5--0.6~$M_{\sun}$ for the pure Fe and Fe+H mixture cases, respectively. For a Type~Ia explosion, we would expect that about half of the ejecta mass is Fe, $\sim0.7$~M$_{\sun}$ \citep[e.g.,][]{Iwamoto1999}. Therefore, our results point to a the clumpy ejecta of a Type~Ia explosion with some admixture of H in the Fe-rich ejecta.

\subsection{\snrb}
\subsubsection{Multi-wavelength morphology}
\label{1234-mwm}
The extent of \snrb\ is best observed in the optical regime, with a clear elliptical shell with \ratio\ consistent with shock excitation. However, as reported in KSP13, the projected elliptical shape is incomplete at the northwestern end with only very faint filamentary structure visible. Comparison of the optical shell to dust emission in the region suggests that \snrb\ has cleared out the dust, with the shell edge corresponding very well to a dust cavity (Fig.~\ref{1234_4panel}-bottom right). 

\begin{figure*}[!t]
\begin{center}
\resizebox{\hsize}{!}{\includegraphics[trim= 0cm 0cm 0cm 0cm, clip=true, angle=0]{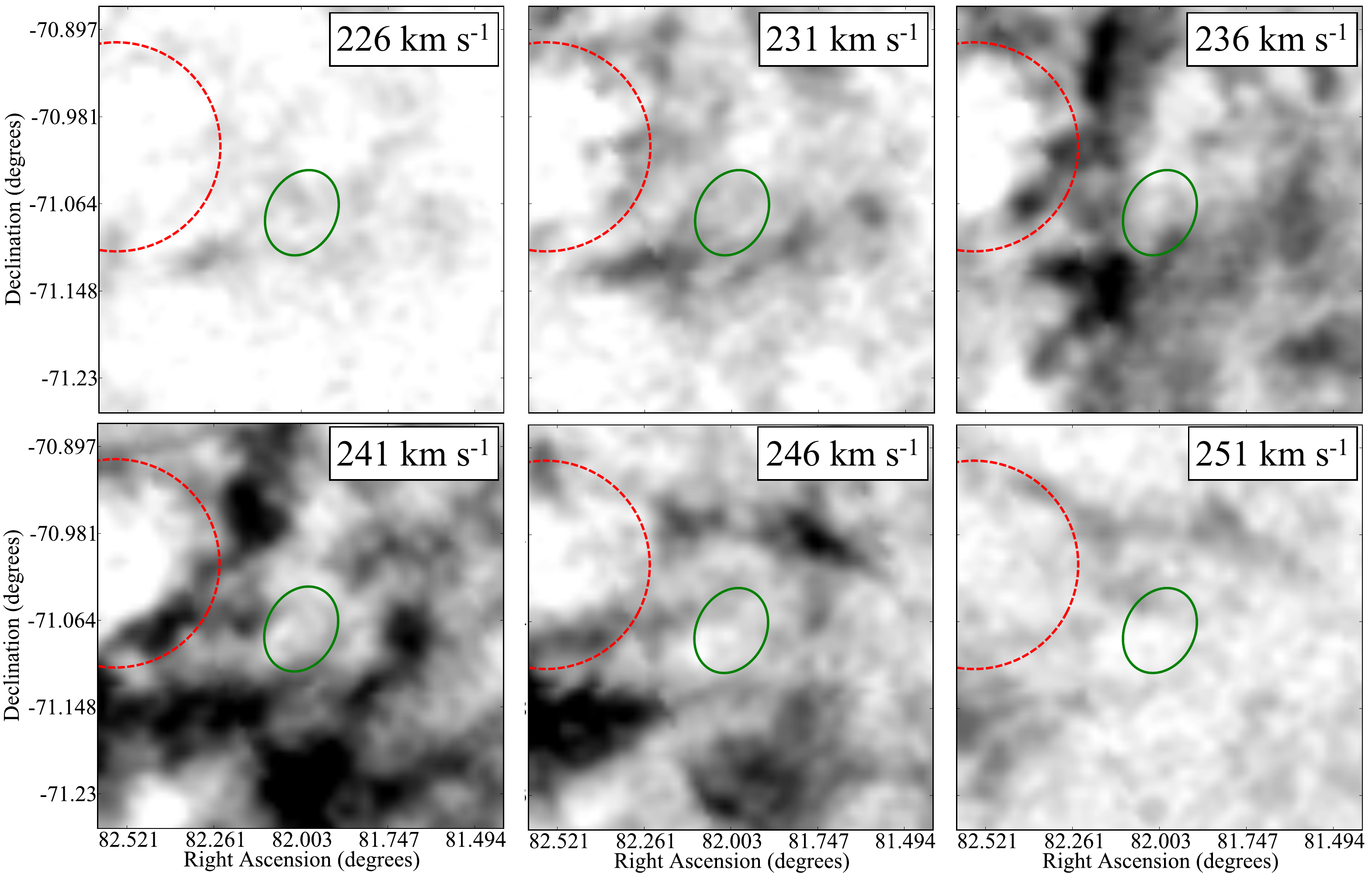}}
\caption{\ion{H}{i} channel maps of the \snrb\ region. The dimensions of the SNR determined in KSP13 are shown by the green ellipse in each panel. The red dashed line mark the approximate position of the giant \ion{H}{i} shell \object{GS~70} \citep{Kim1999} associated with the N~206 star forming region. The heliocentric velocities are also indicated. In the 226~km~s$^{-1}$, 231~km~s$^{-1}$, 236~km~s$^{-1}$, and 241~km~s$^{-1}$ panels there is a noticeable reduction in \ion{H}{i} density to the northwest of the shell.}
\label{1234_HI}
\end{center}
\end{figure*}

The X-ray morphology of \snrb\ is extremely unusual. The emission extends from the southeastern optical shell, right through the southeast-northwest axis of symmetry, appearing to escape or blow-out the northwestern shell (Fig.~\ref{1234_4panel}-bottom left). Blow-outs have been observed at various wavelengths in other LMC remnants, namely \object{N~11L} and \object{N~86} \citep{Williams1999,Crawford2008}. We obtained the MIPS 24~$\mu$m image of \snrb\ (see Section~\ref{ir}), shown in Fig.~\ref{1234_4panel}-bottom right, which indicates a possible reason for the blow-out in the northwest. The dust emission is noticeably absent outside the northwestern shell which might indicate a significantly lower ISM density in the northwest as compared to other regions of the SNR. We investigate this further using data from the \ion{H}{i} survey of the LMC (see Section~\ref{hi}) by extracting channel maps centred on \snrb, shown in Fig.~\ref{1234_HI}. The distribution of \ion{H}{i} in the 226~km~s$^{-1}$, 231~km~s$^{-1}$, 236~km~s$^{-1}$, and 241~km~s$^{-1}$ panels indicates an \ion{H}{i} cavity immediately to the northwest of the SNR. A similar explanation has been proposed for the blow-out from N~11L \citep{Williams1999}. For this reason, we suggest that the blast wave of \snrb\ has blown-out into this low density cavity, allowing the shocked ejecta gas to escape behind it. Given the low density conditions in the ejecta, and that the Fe-rich gas can only lose energy via inefficient radiative cooling, the ejecta can remain X-ray bright as the gas escapes the core into the cavity.

The absence of notable soft X-ray emission within the optical shell with abundances consistent with swept-up ISM could point to either a high absorbing column in the direction of \snrb\, or that the remnant is close to the transition into the radiative phase and the shell has become either too cool to emit X-rays or too faint to be detectable. The low value of $N_{\rm{H,LMC}}$ in the spectral fits (see Table~\ref{1234_tab}) suggests the latter. Moreover, the prominent \sii\ and significant \oiii\ indicates that at least part of the shell is radiative, and some fraction of the gas has cooled down below X-ray emitting temperatures. Other LMC remnants to exhibit similar properties include MCSNR~J0508--6830 and MCSNR~J0511--6759 \citep{Maggi2014}. 

Other members of the evolved Fe-rich SNR class in the LMC also exhibit irregular/elongated cores, though somewhat less pronounced than that of \snrb. We have argued that the unusual core morphology of \snrb\ is the result of a blowout with ejecta escaping into a low density cavity. In other members of the sample, there is no compelling evidence that the irregular/elongated cores resulted from a blowout since the ejecta always appear internal to the optical/X-ray shells. Remnants such as \snra, discussed above, and MCSNR~J0508-6902 \citep{Bozzetto2014} display elongated Fe-rich cores, approximately aligned with the elongated axis of symmetry of the shells, suggesting a common origin. The core of \snrb\ is also elongated along the shell's axis of symmetry and, though the dimensions of the core were likely increased by the blowout, the ellipsoidal shell suggests that some initial elongation could have also been present. Possible external influences which could cause the elongation are ambient density inhomogeneities or ISM magnetic field orientation. Explosion asymmetries offer an intrinsic explanation for the elongation, with Type~Ia explosions thought to be sparked off-set from centre \citep[e.g.,][and references therein]{Maeda2012}. Explosion asymmetries could also explain the Fe-ejecta ``schrapnel'' observed in MCSNR~J0511-6759 outside the bulk Fe core but behind the eastern shell front \citep{Maggi2014}. A truly rigorous treatment of the distribution in the Fe cores of these objects requires deep observations by \chandra\ which offers superior sub-arcsecond resolution. Unfortunately, those remnants with irregular/elongated cores have yet to be observed by \chandra\ with sufficient depth to facilitate such an analysis.

\subsubsection{X-ray emission}
\label{1234-disc}
{\it Ejecta emission:} \\
\noindent The X-ray emission of \snrb\ was found to be entirely ejecta emission with two thermal components representing pure Fe and pure O providing a good fit to the X-ray spectrum.  We found that the emission was dominated by the Fe component with $kT_{\rm{Fe}} = 0.71~(0.69-0.73)$~keV, with a small contribution from O. The temperature of the O component was found to be lower than that of the Fe gas,  $kT_{\rm{O}} = 0.17~(0.13-0.22)$~keV, but consistent with the peak emissivity temperature of the strongest O lines in the 0.3--1~keV range (i.e., $kT = 0.17$~keV). Given this temperature difference, it is unclear whether the Fe and O components are co-spatial. From the SN nucleosynthesis yields of \citet{Iwamoto1999}, the O/Fe ratio is expected to be 0.3--0.7 by number for Type Ia events and $\sim70$ for CC events. Using the determined emission measures of O and Fe, and assuming that they are co-spatial results in an O/Fe ratio of $\sim6-15$, closer to that expected for a Type~Ia explosion but still incompatible with both scenarios. If the O and Fe are not co-spatial then this ratio is not representative of the ejecta. For the purposes of the forthcoming calculations, we assume for simplicity that the Fe and O components are not co-spatial and that Fe dominates the interior volume.

We follow the same procedure outlined in Section~\ref{1139-xray-analysis} for the calculation of the ejecta Fe mass. We fitted an ellipse to the 0.7--1.1~keV extent of the SNR and determined the dimensions of the ejecta emission to be $\sim1\farcm1\times3\farcm0$, corresponding to $30.4~\rm{pc}\times86.6$~pc at the LMC distance. Given the highly elongated morphology of the ejecta emission, it is far more likely that the depth of the emission is closer to 30.4~pc than 86.6~pc. Therefore, we assume that the third dimension of the ellipsoid is 30.4~pc, yielding a total ejecta volume $V$ of $1.2\times10^{60}$~cm$^{3}$.

For our spectral fit results and estimated $V$, a pure heavy element ejecta results in an Fe mass of  1.9--2.1~$M_{\sun}$, about three times the Fe mass expected for a Type~Ia yield. In the case that an equivalent mass of H is mixed into the Fe-rich ejecta, the estimated Fe mass is in the range 0.9--1.0~$M_{\sun}$, slightly higher than expected for a Type~Ia explosion. Again, it is likely that the ejecta are clumpy and the determined masses must be corrected accordingly (see Section~\ref{1139-xray-analysis}), resulting in an estimated Fe mass of 1.2--1.3~$M_{\sun}$ and 0.6--0.7~$M_{\sun}$ for the pure Fe and Fe+H mixture cases, respectively. It is clear from these results that, a clumpy ejecta with an admixture of H of similar mass to the Fe ejecta is required for agreement between our determined Fe mass and that expected from Type~Ia explosive nucleosynthesis yields \citep[e.g.,][]{Iwamoto1999}.



\section{Summary}
\label{summary}
We presented the analysis of new \xmm\ observations of \snra\ and \snrb, and supplemented these with data from the Australian Telescope Compact Array, the Magellanic Cloud Emission Line Survey, \spitzer, and the \ion{H}{i} survey of the LMC. The main findings for each remnant can be summarised as follows:\\

\noindent {\it \snra}
\vspace{-1mm}
\begin{itemize}
\item We estimated the size of the remnant to be $54.0~(\pm5.0)~\rm{pc}\times42.1~(\pm5.0)$~pc, with the major axis rotated $\sim17^{\circ}$ east of north. The shell X-ray emission is highly correlated with the enhanced \sii\ and radio emission with a common knot in the east and extended emission along the western front. We found that this is most likely due to the SNR encountering a denser medium in these regions. 

\item We measured the radio spectral index of \snra\ using integrated flux measurements, a spectral index map, and T-T plots. Although some variation in the determined spectral index was found depending on the method, all the results point to the emission being non-thermal in nature. The most reliable results, the T-T plot method, showed a spectral index consistent with the standard $-0.5$ for SNRs. 

\item Radio polarisation at 6~cm indicate a higher degree of polarisation along the western front and at the eastern knot, with a mean fractional polarisation across the remnant of $P \cong (20 \pm 6)\%$. The field vectors were oriented tangential to the shock front in the west, typical of older remnants where there is compression of the interstellar ambient field. The field vectors in the knot to the east show a somewhat radial orientation.

\item We detected a two-component X-ray emission from \snra\ with a central core dominated by Fe L-shell emission surrounded by a softer shell. The soft shell is typical of swept-up and shock-heated ISM, consistent with an SNR in the Sedov phase, whereas the Fe component is most likely reverse shock-heated ejecta.

\item Using the spectral fit results for the shell and the Sedov self-similar solution, we estimate the age of \snra\ to be $\sim16-28$~kyr, with an initial explosion energy of $(0.07-0.84)\times10^{51}$~erg. We calculated the mass of Fe contained in the Fe-rich core, which is consistent with expected yields from Type~Ia explosive nucleosynthesis models if the ejecta are clumpy with some admixture of H in the Fe.

\end{itemize}

\vspace{1mm}
\noindent {\it \snrb}
\vspace{-1mm}
\begin{itemize}
\item The detected X-ray emission from \snrb\ was also found to be dominated by Fe L-shell emission. However, unlike \snra\, no soft X-ray shell was detected. The X-ray morphology of the Fe-rich ejecta was extremely unusual, extended to the north beyond the apparent optical shell. The 24~$\mu$m and \ion{H}{i} data indicate a low density cavity in the ambient medium to the north of the optical shell and we suggested that the blast wave has broken out into this cavity, allowing the shock heated ejecta to escape.

\item We calculated the mass of Fe contained in the ejecta, which is also consistent with expected yields from Type~Ia explosions if the ejecta are clumpy with some admixture of H in the Fe. 

\end{itemize}

\noindent \snra\ and \snrb\ are new members of the recently identified class of evolved SNRs with centrally-peaked, Fe L-shell dominated emission \citep{Borkowski2006}, bringing the known population of such remnants in the LMC to 11. The overwhelming dominance of Fe emission lines in their spectra is evidence that they result from Type~Ia explosions. The Fe-rich plasmas were found to be in collisional ionisation equilibrium, consistent with other members of the class and suggests that \snra\ and \snrb\ most likely resulted from the `prompt' channel of single degenerate systems. The Fe cores of both remnants, in particular \snrb, we found to be elongated/irregular, which appears to be a relatively common feature of this class of evolved SNRs.

\begin{acknowledgements} We would like to thank our referee Fred Seward for reviewing our manuscript and for his suggestions to improve the paper. This research is funded by the Bundesministerium f\"{u}r Wirtschaft und Technologie/Deutsches Zentrum f\"{u}r Luft- und Raumfahrt (BMWi/DLR) through grant FKZ 50 OR 1309. M.S. acknowledges support by the Deutsche Forschungsgemeinschaft through the Emmy Noether Research Grant SA2131/1-1. P.\,M. acknowledges support by the Centre National d'\'Etudes Spatiales (CNES). E.T.W. acknowledges support by the Deutsche Forschungsgemeinschaft through the Research Grant WH-172/1-1. Cerro Tololo Inter-American Observatory (CTIO) is operated by the Association of Universities for Research in Astronomy Inc. (AURA), under a cooperative agreement with the National Science Foundation (NSF) as part of the National Optical Astronomy Observatories (NOAO). We gratefully acknowledge the support of CTIO and all the assistance that was provided for upgrading the Curtis Schmidt telescope. The MCELS project has been supported in part by NSF grants AST-9540747 and AST-0307613, and through the generous support of the Dean B. McLaughlin Fund at the University of Michigan, a bequest from the family of Dr. Dean B. McLaughlin in memory of his lasting impact on Astronomy. We used the karma software package developed by the ATNF. The Australia Telescope Compact Array is part of the Australia Telescope, which is funded by the Commonwealth of Australia for operation as a National Facility managed by CSIRO. \end{acknowledgements}

\bibliographystyle{aa}

\end{document}